\documentclass[sigconf]{acmart}
\AtBeginDocument{%
  }

\usepackage{pgfplots}
\usepackage{pgfplotstable}
\pgfplotsset{compat=1.7}

\pgfplotsset{every axis/.append style={
                    label style={font=\scriptsize},
                    tick label style={font=\scriptsize}  
                    }}
\usepackage{subcaption}

\usepackage[misc]{ifsym}
\usepackage{bbding}
\usepackage[ruled,linesnumbered]{algorithm2e}
\usepackage[utf8]{inputenc}
\usepackage{graphicx} 
\usepackage{pdflscape}
\usepackage{afterpage}
\usepackage{adjustbox}
\usepackage{enumerate}
\usepackage{xcolor} 
\usepackage{mdframed} 
\usepackage{enumitem}
\usepackage[most]{tcolorbox}
\usepackage{flushend}

\mdfdefinestyle{grayboxstyle}{
    backgroundcolor=gray!20, 
    linewidth=0pt, 
    innerleftmargin=10pt, 
    innerrightmargin=10pt, 
    innertopmargin=10pt, 
    innerbottommargin=10pt 
}

\pgfplotsset{
SmallBarPlot/.style={
    font=\footnotesize,
    ybar,
    width=\linewidth,
    ymin=0,
    xtick=data,
    xticklabel style={text width=1.5cm, rotate=90, align=center}
},
BlueBars/.style={
    fill=blue!20, bar width=0.25
},
RedBars/.style={
    fill=red!20, bar width=0.25
},
DataPlot/.style={
    font=\footnotesize,
    width=\linewidth
}
}

\usepackage{amsmath}    
\usepackage{todonotes} 
\usepackage[acronym,toc]{glossaries} 
\usepackage{comment}

\usepackage{xcolor} 
\usepackage{mdframed} 

\mdfdefinestyle{grayboxstyle}{
    backgroundcolor=gray!20, 
    linewidth=0pt, 
    innerleftmargin=10pt, 
    innerrightmargin=10pt, 
    innertopmargin=10pt, 
    innerbottommargin=10pt 
}

\newboolean{showcomments}
\setboolean{showcomments}{true} 
\ifthenelse{\boolean{showcomments}}
{\newcommand{\nb}[2]{
		\fcolorbox{black}{yellow}{\bfseries\sffamily\scriptsize#1}
		{\sf\small$\blacktriangleright$\textit{#2}$\blacktriangleleft$}
	}
	
}
{\newcommand{\nb}[2]{}
	
}

\usepackage{xspace}

\usepackage{enumitem}

\setcopyright{acmcopyright}
\copyrightyear{2024}
\acmYear{2024}
\acmDOI{XXXXXXX.XXXXXXX}

\acmConference[SEAMS'24]{19th International Conference on Software Engineering for Adaptive and Self-Managing Systems}{April 15--16, 2024}{Lisbon, Portugal}
\acmPrice{15.00}
\acmISBN{978-1-4503-XXXX-X/18/06}

\begin{document}
\title{Exploring the Potential of Large Language Models in Self-adaptive Systems}

\author{Jialong Li}
\affiliation{%
  \institution{Waseda University}
  \city{Tokyo}
  \country{Japan}}
\email{lijialong@fuji.waseda.jp}

\author{Mingyue Zhang}
\affiliation{%
  \institution{Southwest University}
  \city{Chongqing}
  \country{China}}
\email{myzhangswu@swu.edu.cn}

\author{Nianyu Li}
\authornote{Corresponding author: Nianyu Li, li\_nianyu@pku.edu.cn}
\affiliation{%
  \institution{ZGC National Laboratory}
  \city{Beijing}
  \country{China}}
\email{li\_nianyu@pku.edu.cn}

\author{Danny Weyns}
\affiliation{%
  \institution{KU Leuven}
  \city{Leuven}
  \country{Belgium}}
\email{danny.weyns@kuleuven.be}

\author{Zhi Jin}
\affiliation{%
  \institution{Peking University}
  \city{Beijing}
  \country{China}}
\email{zhijin@pku.edu.cn}

\author{Kenji Tei}
\affiliation{%
  \institution{Tokyo Institute of Technology}
  \city{Tokyo}
  \country{Japan}}
\email{tei@c.titech.ac.jp}

\renewcommand{\shortauthors}{J Li. et al.}

\begin{abstract}
Large Language Models (LLMs), with their abilities in knowledge acquisition and reasoning, can potentially enhance the various aspects of Self-adaptive Systems (SAS). Yet, the potential of LLMs in SAS remains largely unexplored and ambiguous, due to the lack of literature from flagship conferences or journals in the field, such as SEAMS and TAAS. The interdisciplinary nature of SAS suggests that drawing and integrating ideas from related fields, such as software engineering and autonomous agents, could unveil innovative research directions for LLMs within SAS. To this end, this paper reports the results of a literature review of studies in relevant fields, summarizes and classifies the studies relevant to SAS, and outlines their potential to specific aspects of SAS.
\end{abstract}


\maketitle

\section{Introduction}
A language model is, in essence, a computational artifact that allows predicting the next word in a sentence based on the probability distribution over sequences of words~\cite{openai2023gpt4}. 
Empowered by modern AI techniques such as the Transformer~\cite{Transformer}, pre-trained language models (PLM)~\cite{devlin2019bert} have demonstrated increasing performance in specific tasks following the 'pre-train and fine-tune' paradigm. Furthermore, with the increasing scale of the model size and the amount of data used for pre-training and fine-tuning, PLMs show emergent and remarkable capabilities in various tasks~\cite{wei2022emergent}. Such large-scale PLMs are commonly referred to as `large language models (LLM)'.

As discussed in the fields of `Cognitive Linguistics' and `Philosophy of Language', language not only serves as an intermediary that enables the building of humans' complex systems of knowledge but also defines a deeper logical structure, reflecting the logic of human thought. Similarly, LLMs, trained on vast textual corpora, have shown their capabilities in knowledge acquisition as well as in logical reasoning and planning. Leveraging these abilities, LLMs have demonstrated strong capabilities in solving diverse problems, leading to a rapid expansion of research and applications. 

Self-adaptive Systems (SAS) are engineered to autonomously adapt to their dynamics in their environment or internal changes without manual interventions, a capability essential for tackling real-world challenges~\cite{MAPE,Cheng2009,1350726,SAS_Intro,10.1145/3589227}. 
LLMs, as demonstrated in other studies, have significantly enhanced the system's capabilities, including context awareness and decision-making, which are essential for handling self-adaptation.
However, there is a notable scarcity of literature on the use of LLM within the field of SAS, particularly from flagship conferences or journals in the field like SEAMS, ACSOS, and TAAS.
This lack of research makes the potential of LLMs in SAS remain unexplored and ambiguous.
As SAS is a cross-disciplinary research field that intersects with software engineering, autonomous agents, human-computer interaction, among others, we believe that a cross-pollination from these relevant fields, could lead to innovative insights, helping to identify potential research directions of LLMs in the context and perspective of SAS.

To this end, this paper aims to explore the potential of LLMs in SAS by targeting the following two research questions:
\begin{itemize}
\item RQ1: What are the key areas of research that explore the application of LLMs relevant to SAS?
\item RQ2: How do the existing studies on LLMs contribute to specific aspects of SAS, and what future research challenges do they suggest for the SAS research community? 
\end{itemize}
To answer RQ1, we review LLM studies from four different research fields relevant to SAS and summarize key research areas on LLM within these fields. 
To answer RQ2, we further filter and categorize the above studies, discussing LLMs' potential and challenges in various aspects of SAS.
We start with a brief introduction of LLM (Section~\ref{sec: background}) and explain the study design (Section~\ref{sec: studyDesign}). Then,  we present the results answering the research questions (Sections~\ref{sec: RQ1} and \ref{sec: RQ2}). Finally, we discuss threats to validity (Section~\ref{sec: ttv}), and draw conclusions (Section~\ref{sec: conclusion}).

\section{Background: Large Language Model}
\label{sec: background}

To lay the foundation for the subsequent discussions, we briefly introduce LLM focusing on core aspects.

\textbf{Architectures of LLMs and Pre-training.}
LLMs refer to transformer-based, large-scale language models that contain billions of parameters and are pre-trained on massive text data~\cite{kaplan2020scaling}. 
For instance, GPT-3 has 175 billion parameters and uses pre-processed 570GB of text data for training. The architecture of LLMs can be categorized into three main types: (i) encoder-only, in which the encoder encodes the input text into a hidden representation to capture the relationships between words and the overall text context; (ii) encoder-decoder, in which the encoder processes the input into a hidden space, and the decoder translates the abstract representation from hidden space into relevant text expression; and (iii) decoder-only, which is used in models like GPT, gradually generates the output text by sequentially predicting the subsequent tokens. 
LLMs are pre-trained using datasets that include web pages, books, conversational text, and program code. The data undergoes pre-processing, such as quality filtering, de-duplication, to improve data quality, and privacy data reduction to enhance privacy.

\textbf{Fine-tuning of LLMs.}
After pre-training, fine-tuning is a technique that uses application-specific customized datasets to provide additional training of the model, thereby improving its performance on the specific task~\cite{LMFineTuning}. A representative example is OpenAI's Codex~\cite{chen2021evaluating} that is based on GPT-3 and specifically fine-tuned for code generation. In addition to traditional fine-tuning, LLMs require two new types of tuning:
(i) Instruction tuning to enhance LLM ability to accurately comprehend and execute tasks as directed by (user-given) natural language instructions; and (ii) Alignment tuning to align LLM more closely with human values such as helpfulness and honesty.

\textbf{Utilizing LLMs.}
After pre-training and optional fine-tuning, LLMs can be utilized to solve various tasks when given suitable prompts. The prompting strategies can be mainly classified into three categories: (i) In-Context Learning (ICL), which offers task description along with examples as demonstrations~\cite{ICL}; (ii) Chain-of-Thought (CoT) that incorporates intermediate steps of reasoning into the prompts~\cite{CoT}; and (iii) Planning that decomposes complex tasks into smaller sub-tasks and creates a plan of actions to complete the overall task~\cite{zhou2023leasttomost}. Employing these strategies enhances LLM performance by aligning the model's processing with the nature of the specific task, thereby resulting in the generation of more coherent and contextually relevant responses. 

\textbf{Capabilities of LLMs.}
According to \citet{LLM_survey}, LLMs mainly have the following six abilities:
(i) Language generation aims at creating text that meets certain requirements set by tasks like translation, summarization, question answering, as well as formal language (e.g., logical forms, code) synthesis;
(ii) Knowledge utilization entails leveraging the extensive factual knowledge from the pre-training corpus or accessing external data to perform tasks that require significant knowledge, such as common-sense question answering and fact completion;
(iii) Reasoning, encompassing knowledge, symbolic, and mathematical reasoning, pertains to the capability to comprehend and apply evidence or logic to reach conclusions or make decisions;
(iv) Human alignment means that LLM can well conform to human values and needs, with criteria like truthfulness, honesty, and safety;
(v) Interaction with the external environment refers to the ability to receive feedback from the external environment and perform actions according to behavioral instruction, e.g., generating detailed action plans in natural language or other formats;
(vi) Tool manipulation refers to the ability to utilize external tools, such as search engines and calculators, to enhance performance on given tasks.
It should be noted that these six capabilities are neither comprehensive nor orthogonal, i.e., a task may require a combination of multiple capabilities.

\textbf{Shortcomings of LLMs.}
Although this paper does not delve into the technical details of LLMs, it is important to consider shortcomings when using them. We highlight key issues. 
First, hallucination refers to the phenomenon where a model generates misleading and factually incorrect information. This may challenge the system's reliability and trustworthiness.
Second, the need for high-performance hardware: Due to the sheer model size and complex inference, LLMs generally require high-performance hardware.
This leads to increased deployment and operational costs;
Third, slow inference speed: Similarly, due to the large model size, LLMs take time to process input and generate output. This limits the application of LLMs in domains requiring real-time responses.
Fourth, explainability issues: Due to LLMs' complex neural network architectures and massive parameter counts, they often operate as black boxes. This may make it difficult to apply them to critical applications.
Fifth, privacy concerns:  LLMs may use and sometimes share user data with third parties (e.g., when using external search engines). This may not align with users’ data privacy expectations. 
These issues call for further research and regulation. 

\section{Study Design}
\label{sec: studyDesign}
We used a study design comprising five steps.
The first two steps focused on searching for literature on LLMs in fields relevant to SAS (targeting RQ1).
The last three steps focused on filtering and classifying the literature in relation to the field of SAS (targeting RQ2). 

Step 1: Due to the cross-disciplinary nature of the SAS area, we collected studies from fields relevant to SAS. Specifically, we selected two theoretical research fields (software engineering and autonomous agents) and two application-oriented domains (robotics and autonomous driving). 
The choice of these fields is motivated by their relevance to adaptation and the abundance of literature in these areas.
To more effectively search for papers and avoid reinventing the wheel, we collected literature from well-known reviews in these fields. Specifically, we selected a total of seven survey papers: two for software engineering~\cite{LLMonSE_Survey1,LLM4SE_survey2}, two for autonomous agents~\cite{LLM4agent_survey, LLM4agent_2}, one for robotics~\cite{LLM4robotics}, and one for autonomous driving~\cite{LLM4AD_survey}. In addition, we also selected a review paper for LLM~\cite{LLM_survey} to capture potential literature beyond the aforementioned fields.
After de-duplication, this resulted in a total of 1,363 papers. 

Step 2: Preliminary screening of the literature based on the following criteria: (i) Exclude papers published before 2017, as the Transformer model was introduced in that year; (ii) Exclude GitHub projects, news, X (Twitter), and blogs; and (iii) Exclude papers unrelated to PLM or LLM by checking the title and abstract of studies. 
This resulted in 880 papers after the preliminary screening.

Step 3: Establish first-level categories for classifying studies. 
The first-level categories followed the MAPE-K reference model (Monitor-Analysis-Plan-Execution over Knowledge), the classic conceptual model for engineering SAS~\cite{MAPE}. Additionally, we included the category of `human-in-the-loop (HITL).'

Step 4: Confirm the relevance to SAS of each study. The relevance was determined by whether a study explicitly contributes to the first-level categories (from step 3), i.e., whether LLMs empower context awareness, decision-making, plan execution, and human-computer interaction with the system. To enhance the validity of this step, two of the researchers involved in this study independently confirmed the title, abstract, and keywords of each study and decisions were made based on discussions between the two in case of disagreement.
We mainly excluded theoretical studies of LLM (e.g., fine-tuning techniques), studies that improve the performance of LLM without direct relevance for SAS (e.g., prompt engineering), applications of the humanities, and dataset or benchmark studies. This step resulted in 179 papers.

Step 5: Refine categories and classify the selected studies. We thoroughly read the full text of all selected studies and categorized them into the first-level categories. We then refined each first-level category into subsequent-level categories and classified each study. This step involved at least three researchers of the study.

\section{RQ1: LLM in Relevant Fields}
\label{sec: RQ1}
In this section, we investigate the application of LLMs in fields relevant to SAS. The results enable us to answer RQ1.

\textbf{Software Engineering.}
LLMs are instrumental across various stages of Software Engineering (SE), offering tools and methods that enhance the efficiency and quality of SE activities~\cite{LLMonSE_Survey1,LLM4SE_survey2}.
In requirements engineering, LLMs are used for classifying, analyzing requirements~\cite{LLM_goalmodel}, and translating requirements in natural language to logic-based specifications~\cite{LLM_req_translation}.
During the software design phase, LLMs facilitate the generation of software prototypes~\cite{LLM_prototyping}.
In the development phase, their applications include code generation and completion, code comprehension and summarization~\cite{ma2023chatgpt}, and recommending or utilizing APIs (Application Programming Interface)~\cite{qin2023toolllm}.
LLMs are also explored for software quality assurance and maintenance. They are mainly applied to automated test generation~\cite{LLM_test_generation} , program repair~\cite{LLM_program_repair}, and code review~\cite{LLM_code_review}.

\textbf{Autonomous Agents.}
LLMs have already been used to support various types of autonomous agents, including simulation agents, web agents, game agents, assistant agents, and more~\cite{LLM4agent_survey, LLM4agent_2}.
Although there is no consensus on the architectural design of LLM-based autonomous agents, LLMs have been primarily used in the following five kinds of modules:
(i) Profiling module, where LLMs are utilized to automatically define the appropriate agent's roles, such as coders or domain experts, for given tasks~\cite{chen2023agentverse};
(ii) Perception module, where multi-modal LLMs enable the perception of various modalities (textual, visual, and auditory)~\cite{lyu2023macawllm} and formats (e.g., point cloud-based 3D maps)~\cite{graule2023ggllm};
(iii) Memory module, in which LLMs are used to store, retrieve, or reflect past information, utilizing short- and/or long-term memory structures in various formats like natural languages, embedding vectors, and databases~\cite{zhu2023ghost};
(iv) Decision-making module (planning and reasoning), with representative methods like Chain of Thought (CoT) and self-consistency. In addition, various reflection methods are also proposed, enabling LLMs to modify and refine agent plans based on feedback from humans or the environment~\cite{wang2023describe};
(v) Action module, in which LLMs are used to decide actions based on the given decision, usually by calling external tools such as APIs, planners, and solvers~\cite{zhang2023toolcoder, qin2023tool}

\textbf{Robotics.}
Robotics can be viewed as embodied agents that solve their own domain-specific problems~\cite{LLM4robotics}.
For instance, \cite{TaPA} aligns LLMs with visual perception models, enabling the generation of executable plans based on the perceived objects in the scene.
\cite{SayPlan} allows LLMs to conduct efficient semantic search-based planning based on the specific input of 3D scene graphs.
Additionally, there are specialized pre-trained Vision-Language-Action models (VLA) for robotics, such as Google's PaLM-E and Deepmind's RT-2~\cite{Deepmind_RT2}.

\textbf{Autonomous Driving.}
LLMs are extensively studied in the domain of autonomous driving. In perception, LLMs are used to assist in tracking or detecting 2D or 3D objects in driving scenes~\cite{ding2023hilmd}. For decision-making, LLMs have been explored for tasks such as generating motion plans by transforming these tasks into sequence modeling problems~\cite{driving_with_LLM}.
Additionally, another promising direction involves scene generation, where LLMs combined with diffusion multimodal models are utilized to generate realistic driving videos or intricate driving scenarios under specific environmental factors~\cite{GATA1}. 
This approach significantly lowers the cost of data collection and labeling, thereby providing an efficient and cost-effective solution for the verification and testing of autonomous driving systems.

\begin{tcolorbox}[breakable, colback=gray!20, colframe=gray!100, arc=0.3mm,   leftrule=3pt, rightrule=0pt, toprule=0pt, bottomrule=0pt,   left=1pt, right=1pt, top=2pt, bottom=2pt]
\textbf{Answer to RQ1}
The key areas of research on using LLMs in Software Engineering (SE) lie in improving the efficiency and performance of SE activities. Hence, such improvements offer potential benefits to the engineering of SAS.
Key areas of research on using LLMs in autonomous agents, robotics, and autonomous driving focus on the collection and representation of knowledge and decision-making. These areas offer potential for supporting the runtime activities of SAS, in particular analysis and planning.
\end{tcolorbox}
\section{RQ2: Potential of LLM for SAS}
\label{sec: RQ2}
We now zoom in on the analysis of the potential of LLM for SAS. The results enable answering RQ2. 
We begin with an overview of the classification of the literature. Then, we delve into the potential of LLM for SAS in the different categories.

\subsection{Overview of Literature Classification}
\begin{figure}[th!]
\vspace{-0.1cm}
\includegraphics[width=1\linewidth]{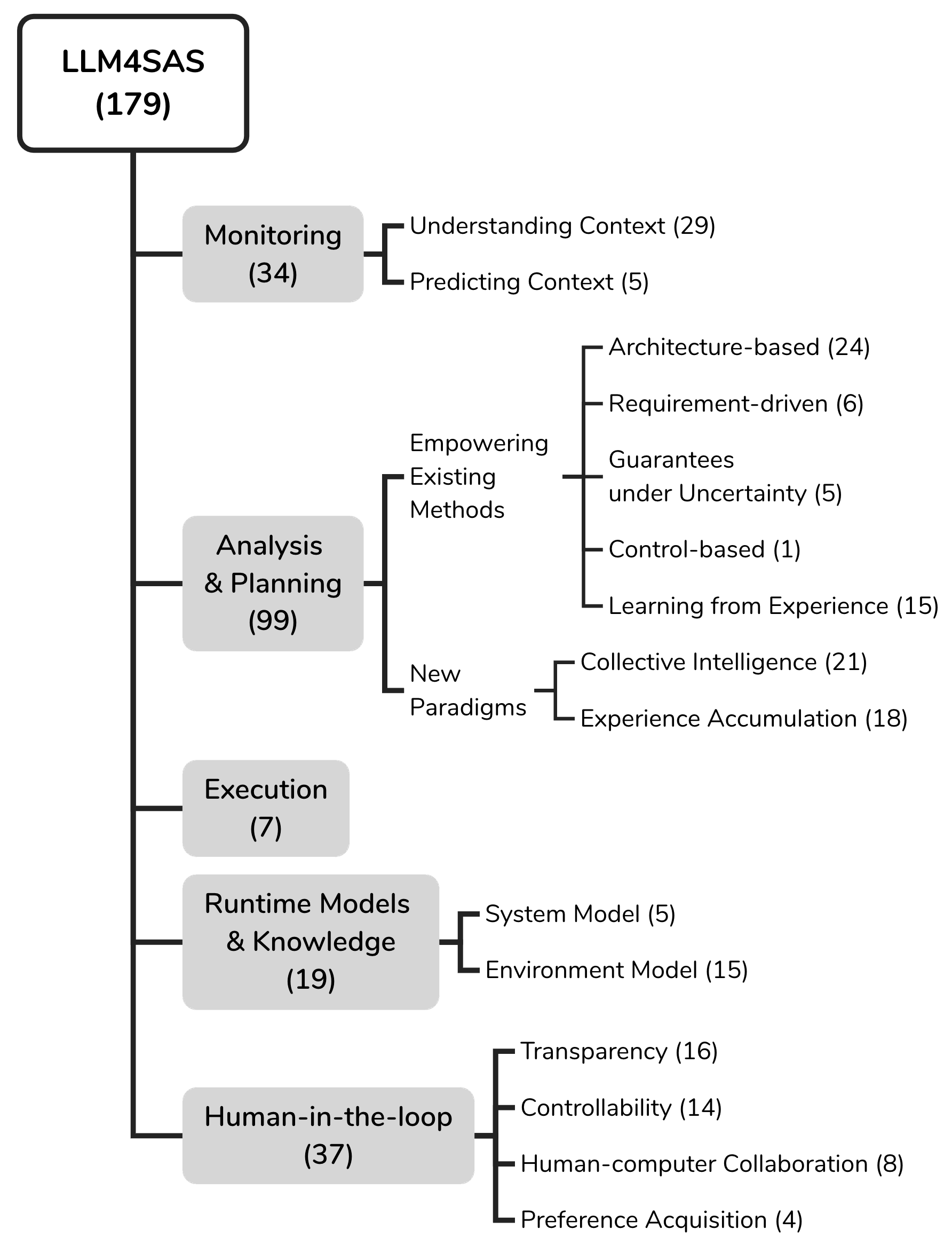}
\caption{Overview of Categorization. 
The numbers in brackets indicate the number of studies in the category. One study can be categorized into multiple categories.
}
\label{classfication}
\vspace{-0.2cm}
\end{figure}

Following the procedure explained in Sec.~\ref{sec: studyDesign}, we classified the collected literature according to the functions of the MAPE-K reference model plus human-in-the-loop, as shown in Fig.~\ref{classfication}.

For the classification, several points require special attention: (i) We combined analysis and planning into a single category, as their responsibilities often overlap and are difficult to clearly distinguish;
(ii) Studies do not necessarily belong to one category; e.g., task assignment for human-computer collaboration could be considered in both planning and human-in-the-loop. We classified each study according to its main category;  
(iii) Due to page limitations, we only discuss representative studies for each classification.
For studies with relatively close contributions or novelty, we select the one or two most relevant based on discussions among the authors. 
The complete literature classification\footnote{Anonymous link for blind review: \url{https://figshare.com/s/09ff1b1c1ed433d55a53}} is available for interested readers.

The remainder of this section presents the potential of LLM in each category and provides an answer to RQ2 for each category.

\subsection{Monitoring}
As SASs operate in environments that are often dynamic and unpredictable, the monitoring function is crucial for collecting and filtering data, as well as understanding the context of the system and the environments in which it operates.

\textbf{Understanding Context}.
In other research fields, language models for vision are a highly discussed topic, yet most SASs do not involve image processing.
However, LLMs also offer interesting possibilities in understanding text or data.
For instance, LLMs can detect semantic anomalies in data~\cite{LM_finance}.
A straightforward example is an autonomous driving system mistaking a truck's taillight for a moving traffic signal.
This ability can not only identify sensing errors or data fusion errors but can also be used to timely and accurately trigger the system's adaptation or human intervention.

\textbf{Predicting Context}.
In addition to understanding context, LLMs can also be used to reason about and predict future contexts. 
\citet{seff2023motionlm} utilizes LLMs to predict the motion of autonomous vehicles by representing continuous trajectories as sequences of discrete motion tokens and defining motion prediction as a language modeling task. 
\citet{graule2023ggllm} uses a semantic map of the environment to infer probable sequences of human actions and activities.
The capability of context prediction allows SAS to anticipate future states or changes, thereby laying a foundation for proactive adaptation.

\begin{tcolorbox}[breakable, colback=gray!20, colframe=gray!100, arc=0.3mm,   leftrule=3pt, rightrule=0pt, toprule=0pt, bottomrule=0pt,   left=1pt, right=1pt, top=2pt, bottom=2pt]
\textbf{Answer to RQ2: Monitoring}

LLMs have demonstrated the ability to enable more accurate and richer contextual awareness, laying a solid basis for higher-quality decision-making in SAS. 
However, the strategies leveraging these abilities into SAS require exploration, as massive sources of (indirect) data will lead to a large design space for monitoring. Furthermore, employing LLM for monitoring tasks in large-scale self-adaptive systems raises performance challenges when high-frequency monitoring is required during runtime. 
\end{tcolorbox}

\subsection{Analysis \& Planning}
\label{sec: analyzer_planner}
The analysis and planning functions provide the mechanisms to analyze the options for adaptation and compose a plan for adaptation to achieve the system's goals.
Here, we divide the literature in two categories: first, using LLM to empower existing approaches, leveraging the `seven waves' in the textbook~\cite{SAS_Intro},\footnote{We omit `automating tasks', which refers mainly to MAPE-K as a whole, and `runtime models', which we discuss as part of ·runtime models \& knowledge'.} and second,  emerging analysis and planning paradigms enabled by LLM. 

\subsubsection{Empowering Existing Methods} \mbox{\ }\vspace{5pt}\

\textbf{Architecture-based Adaptation}. The focus is on architectural models and modeling abstractions that enable the system to reason about adaptation decisions~\cite{1350726}.
As extensively discussed in the literature, LLMs can decompose given tasks (equivalent to the goal management layer in the three-layer model~\cite{3_layer_model}).
Based on the decomposed sub-goals, LLMs can select and use the appropriate tools, which can be considered system components, at runtime (equivalent to the change management layer and component control layer).
For instance, \citet{qin2023toolllm} utilizes ChatGPT to generate a chain of API calls for the given instruction from pools of more than 16,000 real-world APIs.
\citet{zohar2023lovm} uses LLMs to choose appropriate AI models for given tasks, according to the function descriptions of each model, and connect them to complete the task.
Hence, LLMs have the potential to support flexible and large-scale architecture-based adaptations utilizing large-scale libraries of components.

\textbf{Requirements-driven Adaptation}.
This approach considers requirements as first-class entities to manage requirement-driven adaptation.
A recognized open challenge in this approach is how to support automatically changing goals at runtime~\cite{SAS_Intro} to handle the uncertainty and dynamics of requirements.
First, as mentioned in RQ1, many LLM studies in SE have focused on the automated decomposition, analysis, and classification of `static' requirements; thus, bridging them into runtime could be a possible direction.
Second, some studies on autonomous agents also provide insightful ideas for this approach.
\cite{abdelnabi2023llmdeliberation} applies LLM-based agents to negotiation games, which could be a solution for resolving requirement conflicts or enabling requirement relaxation.
\cite{song2023selfrefined} proposes a self-refinement mechanism for automating the design of a reward function, where adjusting the reward function can be viewed as a modification to the requirements and their weights or priorities.

\textbf{Guarantees under Uncertainty}.
This approach aims to provide evidence that the adaptation behavior guarantees the adaptation goals under uncertainty.
It is noteworthy that we did not find work combining LLM with formal methods, such as probabilistic or statistical verification.
However, there are studies based on game theory that offer potential, as game theory can be used to model the behavioral uncertainty of cooperative and adversarial entities and guarantee the system's performance under the worst possible scenarios.
For instance, Alympics~\cite{mao2023alympics} employs an LLM-based agent to simulate multi-agent game simulations, where the LLM enables the construction of more realistic and dynamic models of human interactions.
However, to our knowledge, existing studies have only demonstrated LLMs' ability to generate agent behaviors for defining and simulating more realistic and complex game processes.
To our knowledge, the potential of LLMs in solving mathematically defined games has not yet been discussed.

\textbf{Control-based Software Adaptation}.
This approach focuses on exploiting the mathematical basis of control theory for designing SASs and guaranteeing key properties of the system.
As mentioned in \cite{SAS_Intro}, the key challenge of this approach is how to translate non-functional requirements (NFR) in the context of SE into control properties in the context of control theory (and vice versa).
To our knowledge, although \cite{LLM_adaptivePID} involves using LLMs to interpret the behavior of the PID controller, there are currently no studies using LLMs to automate or assist in building controllers based on requirements.
However, given LLMs' understanding of NFRs, as well as their capabilities in mathematical reasoning and translation, we believe that it is worth exploring the use of LLMs in bridging SE and control theory and apply this to the field of self-adaptation.

\textbf{Learning from Experience}.
This approach is about exploiting machine learning techniques to support functions within managing systems, addressing the growing system scale and increasingly complex uncertainty.
Here, LLM may have several potential applications.
Firstly, LLMs themselves can be used for context-aware classification, such as failure mode classification~\cite{stewart2023FMC}, to reduce the decision-making space to only relevant adaptation options.
Secondly, LLMs have also been applied in automated machine learning and data science to reduce manual design costs and improve performance. 
For example, \cite{LLM_feature_enginering} uses LLMs to iteratively generate semantically meaningful and domain-specific features to automate feature engineering in machine learning.
Thirdly, LLMs have been effectively integrated with reinforcement learning (RL) in several ways.
\cite{song2023selfrefined} proposes an LLM-based self-refinement mechanism for automated reward function design, demonstrating its effectiveness in dynamic environments and complex tasks in robotics.
\cite{xu2023Werewolf} proposes a combined decision-making method of RL and LLM for the Werewolf game. Specifically, LLMs are first used to generate possible action candidates (by applying deductive reasoning to analyze the hidden roles of other players), and RL is subsequently applied to select the optimal action.

\subsubsection{New Paradigms}
\mbox{\ }\vspace{5pt}\

\textbf{Collective intelligence.} Collective intelligence in the form of crowd-sourcing has emerged as a novel method to leverage the wisdom of crowds (where multiple agents often play different roles or are enabled by multiple fine-tuned LLMs) to achieve consensus-driven decision-making through discussion, debate, or voting.
\cite{du2023improving} proposes a method to arrive at a common final answer for reasoning problems through multiple rounds of debate, demonstrating the potential of crowd-sourcing in reducing fallacious answers and hallucinations.
\cite{chen2023agentverse} explores automated expert recruitment (deciding what kind of domain expert is needed for the task and then generating their persona) and different forms of crowd-sourcing (democratic or hierarchical).
\cite{zhang2023controlling} transplants the Actor-Critic framework from RL into LLM multi-agent crowd-sourcing, demonstrating its potential to reduce hallucinations and lower usage costs (in terms of token length).
Due to its nature, this paradigm provides a new avenue for decision-making, particularly suitable for decentralized SAS~\cite{DSAS_pattern}.

\textbf{Experience Accumulation}.
Experience accumulation, distinct from existing methods such as machine learning-based classification or reinforcement learning, refers to the ability of an agent to employ LLMs to accumulate experience and gradually learn to plan through continuous trial-and-error.
For failed experiences, the reasons for failure, analyzed by an LLM or provided by a human, can be reflected upon and incorporated into the next round of planning~\cite{wang2023describe}.
Similarly, for successful experiences, LLMs are used to store them in memory or skill pool, and then retrieve and reuse them when encountering similar situations in the future~\cite{zhu2023ghost}.
The paradigm of experience accumulation aligns with the widely acknowledged yet not extensively explored concept of `self-learning SAS'~\cite{zmy_xuebao20}. This paradigm suggests that LLM provides a foundation that enables SAS to learn from collected experiences and autonomously evolve their own ‘learning’ to handle unexpected conditions.

\begin{tcolorbox}[breakable, colback=gray!20, colframe=gray!100, arc=0.3mm,   leftrule=3pt, rightrule=0pt, toprule=0pt, bottomrule=0pt,   left=1pt, right=1pt, top=2pt, bottom=2pt]
\textbf{Answer to RQ2: Analysis \& Planning}

(1) Existing analysis and planning methods may benefit from integrating LLM helping to improve runtime performance or reduce design costs. 
The integration of LLM with established disciplines like formal verification or control theory remains unexplored, despite the probabilistic nature of LLM that aligns well with many of the frameworks of these traditional methods.
(2) Several new generalized analysis and planning paradigms, previously unexplored in the SAS community, have emerged. These paradigms hold great potential, particularly for decentralized SAS and self-learning SAS.
\end{tcolorbox}

\subsection{Execution}
Execution refers to the enactment of actions of the adaptation plan in porder to adapt a SAS, but its scope varies depending on the concrete designs and division of responsibilities between the managed system (i.e., the system that is subject to adaptation) and the managing system (i.e., the MAPE loop that adapts the managed system). 
For instance, consider a plan `moving to the destination.' Execution can vary: (i) the executor simply passes destination coordinates to the managed system that autonomously completes the movement, or (ii) the executor translates the given plan into a more detailed path or even control parameters for the managed system.

For cases such as the first scenario, execution is straightforward and LLMs offer little benefit. 
However, for more complex tasks like the second scenario, as demonstrated by the Vision-Language-Action model (VLA) or 'robotic transformer', LLMs have demonstrated to be beneficial in translating high-level instructions into specific control parameters \cite{Deepmind_RT2}.
Although the literature in this category is quite limited, LLMs essentially have the ability to reason about and convert high-level activities into low-level configuration actions. Despite these capabilities, current SAS literature suggests that execution is often carried out with simple mappings or rule sets. This identifies a gap in understanding which scenarios can truly benefit from LLM-enabled complex conversions.

\begin{tcolorbox}[breakable, colback=gray!20, colframe=gray!100, arc=0.3mm,   leftrule=3pt, rightrule=0pt, toprule=0pt, bottomrule=0pt,   left=1pt, right=1pt, top=2pt, bottom=2pt]
\textbf{Answer to RQ2: Execution}

LLMs have demonstrated the ability to convert higher-level activities into low-level control actions. SAS may benefit from these capabilities.  However, the potential of LLM for the execution of adaptation plans remains an area for further exploration.

\end{tcolorbox}

\subsection{Runtime Models \& Knowledge}
Knowledge or runtime models, extending the notion of development-time models to runtime, are elements that are associated with the four functions of MAPE, enabling runtime reasoning and decision-making.
In some cases, these runtime models are defined with a specific focus, using meticulously designed Domain-Specific Modeling Languages (DSML)~\cite{SAS_Intro}. We focus here on system models and environment models.

\textbf{System Models}.
LLMs have been practically used to generate models describing different aspects of software, including but not limited to requirement models~\cite{LLM_goalmodel}, and architecture models~\cite{metaGPT}.
In addition, LLMs have demonstrated their ability to generate or translate into domain-specific models.
For instance, \citet{LLM_specification_synthesis} converts requirements from natural language to DSML, although the format and operators of/within the DSML need to be specified by the user in the prompt.
\citet{liu2023chipnemo} applies domain-adapted fine-tuning to enable a 5x reduction in LLM model size and better performance in tasks of industrial chip design.
However, the models generated by these studies are often static at development time and do not consider reasoning during runtime.
Therefore, utilizing LLMs to design DSML for runtime models (which aspects or information to capture and how to support decision-making) may still be a research direction that requires further exploration.

\textbf{Environment Models}.
LLMs as world models~\cite{hao2023reasoning}\footnote{A concept in AI referring to a computational model that simulates the real-world environment, behaviors, and dynamics.}, or more broadly, Artificial Intelligence Generated Content (AIGC), have demonstrated their capability in generating environmental data, thereby simulating environmental uncertainties for training or testing SAS.
For instance, \citet{GATA1} uses video, text, and action inputs to generate realistic driving scenarios with fine-grained scene features.
However, addressing the hallucinations of LLMs and aligning simulated environments with the real world remain as important future research directions.

\begin{tcolorbox}[breakable, colback=gray!20, colframe=gray!100, arc=0.3mm,   leftrule=3pt, rightrule=0pt, toprule=0pt, bottomrule=0pt,   left=1pt, right=1pt, top=2pt, bottom=2pt]
\textbf{Answer to RQ2: Runtime Models \& Knowledge}

LLMs can automate or assist in the design or translation of models. Here, the generated objects include not only models describing the system (to support runtime decision-making) but also models of the operational environment (to enable simulation or testing). 
However, the challenges of creating system models that support reasoning and generating environment models that align with the real world continue to be crucial areas for future research.\end{tcolorbox}

\subsection{Human-in-the-Loop}
\label{sec: hci}
Human-in-the-Loop (HITL) ensures that human intelligence and judgment can assist or guide the system’s adaptive behavior, particularly at critical moments, which is vital for enhancing the overall effectiveness and trustworthiness of the system~\cite{HITL_CPS}.

\textbf{Transparency}.
Transparency refers to the degree to which the system's operations and decision-making processes are understandable to the human user~\cite{nianyu_seams20}.
Many studies have provided insights into improving system interpretability, including:
(i) code summarization and comprehension, where LLMs can directly interpret code and document it, thereby explaining the system's operational logic \cite{ma2023chatgpt};
(ii) model explanation, where LLMs can explain decision-making in models such as PID controller models~\cite{LLM_adaptivePID};
(iii) data explanation, which involves explaining the system's operations through log/diagnostic information of systems~\cite{tagliabue2023real}.
Additionally, the capabilities of Q\&A-style interaction \cite{LLM_AD_QA1} and data visualization \cite{Chat2VIS} of LLMs can also provide directions for more user-friendly and intuitive explanations.
With transparency empowered by LLMs, it becomes easier to establish user trust in SAS's behavior and enable informed decision-making by SAS in co-operation with the user.

\textbf{Controllability}.
Controllability refers to the ability of the human user to intervene, guide, or alter the system's operations or decision-making process.
Some literature explores manually correcting the composition of the Chain of Thought (CoT) and LLM-generated workflows, or allowing humans to modify the intermediate outcomes produced by various steps in the workflow~\cite{AIChains}.
Such features have the potential to enhance the safety and reliability of SAS, as they allow humans to intervene in situations where the system may not act normally or optimally.

\textbf{Human-computer Collaboration}.
Human-computer collaboration refers to the cooperative interaction between humans and the system, thereby harnessing the respective strengths of humans and computers for optimal utility~\cite{MAPEK_HMTeaming}.
\citet{gong2023mindagent} proposes MindAgent, a framework that uses LLMs to generate action patterns for non-player character (NPC) collaborators (in a game environment) based on the behavior (history) of human players.
\citet{kannan2023smartllm} apply LLMs to convert high-level task instructions provided as input into a multi-agent task plan, including task decomposition, agent grouping formation, and task allocation.
These approaches demonstrate the potential of LLMs for designing collaborations in SAS that are both user-centric and computer-mediated.

\textbf{Preference Acquisition}.
Preference acquisition refers to the process of gathering and understanding the preferences and needs of users~\cite{seams23_preference}.
\citet{TidyBot} focuses on preferences for the placement of objects in robotic cleaning tasks, where LLMs are applied to summarize and infer generalized and broadly applicable user preferences from just a few interaction examples.
\citet{thomas2023large} uses feedback from real users to predict whether a search result is valuable to the searcher (as a relevance label), and the evaluation deployed on Bing demonstrated that the accuracy is comparable to human labelers.
As another form of context understanding and prediction, LLMs have shown their potential to acquire human preferences, thereby facilitating personalized and user-centric adaptations for enhancing user satisfaction in SAS.

\begin{tcolorbox}[breakable, colback=gray!20, colframe=gray!100, arc=0.3mm,   leftrule=3pt, rightrule=0pt, toprule=0pt, bottomrule=0pt,   left=1pt, right=1pt, top=2pt, bottom=2pt]
\textbf{Answer to RQ2: Human-in-the-loop}

Although HITL is recognized as an important research direction in the field of SAS, the studies in this area are still limited.
With the emergence of LLMs and their demonstrated capabilities and powerfull generalizability in HITL tasks, they could bring significant opportunities in this direction to SAS.
\end{tcolorbox}

\section{Threats to Validity}\label{sec: ttv}
Our study is exploratory in nature and subject to a number of threats to validity.
First and foremost, we selected studies from a set of existing literature reviews.
Evidently, we may have missed relevant papers.
To limit this threat, we selected well-known and multiple reviews, but we acknowledge that a full systematic literature review would better mitigate this threat.
Second, we established a number of rules to determine the relevance of the studies to SAS.
This set of rules is inherently limited, and we acknowledge that we may have missed interesting opportunities reported in the literature.
By leveraging the MAPE-K reference models, and established approaches documented in the literature~\cite{SAS_Intro}, we have tried to minimize this validity threat.
Thirdly, the interpretation of categories for classification may not have been fully objective. To mitigate this potential validity threat, the data was collected by multiple researchers and discussed in cases of disagreement. Furthermore, the results were cross-checked by the other researchers involved in this study.
\section{Conclusion and Future Work}
\label{sec: conclusion}
In this paper, we reviewed recent studies of LLM in the relevant fields of SAS, and categorized their contributions into specific aspects of SAS.
The results primarily suggest that LLMs have the potential to augment existing methods and introduce new opportunities for functionalities. These potentials encompass all phases of the MAPE-K loop and human-in-the-loop.

In future research, we plan to expand this study. More specifically, due to the emphasis on the timeliness of research, we simplified the steps in our literature search. Similarly, because of space limitations, this paper remains necessarily brief regarding most of the literature. Therefore, we plan to conduct a systematic literature review and perform a comprehensive and finer-grained analysis and discussion.

Nevertheless, we hope that this paper provides a useful and comprehensive snapshot of the potential of LLM for the SAS community, thereby facilitating research on LLM applied in SAS.

\bibliographystyle{ACM-Reference-Format}
\bibliography{main}


\begin{thebibliography}{74}


\ifx \showCODEN    \undefined \def \showCODEN     #1{\unskip}     \fi
\ifx \showDOI      \undefined \def \showDOI       #1{#1}\fi
\ifx \showISBNx    \undefined \def \showISBNx     #1{\unskip}     \fi
\ifx \showISBNxiii \undefined \def \showISBNxiii  #1{\unskip}     \fi
\ifx \showISSN     \undefined \def \showISSN      #1{\unskip}     \fi
\ifx \showLCCN     \undefined \def \showLCCN      #1{\unskip}     \fi
\ifx \shownote     \undefined \def \shownote      #1{#1}          \fi
\ifx \showarticletitle \undefined \def \showarticletitle #1{#1}   \fi
\ifx \showURL      \undefined \def \showURL       {\relax}        \fi
\providecommand\bibfield[2]{#2}
\providecommand\bibinfo[2]{#2}
\providecommand\natexlab[1]{#1}
\providecommand\showeprint[2][]{arXiv:#2}

\bibitem[Abdelnabi et~al\mbox{.}(2023)]%
        {abdelnabi2023llmdeliberation}
\bibfield{author}{\bibinfo{person}{Sahar Abdelnabi}, \bibinfo{person}{Amr Gomaa}, \bibinfo{person}{Sarath Sivaprasad}, \bibinfo{person}{Lea Schönherr}, {and} \bibinfo{person}{Mario Fritz}.} \bibinfo{year}{2023}\natexlab{}.
\newblock \bibinfo{title}{LLM-Deliberation: Evaluating LLMs with Interactive Multi-Agent Negotiation Games}.
\newblock
\newblock
\showeprint[arxiv]{2309.17234}~[cs.CL]


\bibitem[Brohan et~al\mbox{.}(2023)]%
        {Deepmind_RT2}
\bibfield{author}{\bibinfo{person}{Anthony Brohan}, \bibinfo{person}{Noah Brown}, \bibinfo{person}{Justice Carbajal}, \bibinfo{person}{Yevgen Chebotar}, \bibinfo{person}{Xi Chen}, \bibinfo{person}{Krzysztof Choromanski}, \bibinfo{person}{Tianli Ding}, \bibinfo{person}{Danny Driess}, \bibinfo{person}{Avinava Dubey}, \bibinfo{person}{Chelsea Finn}, \bibinfo{person}{Pete Florence}, \bibinfo{person}{Chuyuan Fu}, \bibinfo{person}{Montse~Gonzalez Arenas}, \bibinfo{person}{Keerthana Gopalakrishnan}, \bibinfo{person}{Kehang Han}, \bibinfo{person}{Karol Hausman}, \bibinfo{person}{Alexander Herzog}, \bibinfo{person}{Jasmine Hsu}, \bibinfo{person}{Brian Ichter}, \bibinfo{person}{Alex Irpan}, \bibinfo{person}{Nikhil Joshi}, \bibinfo{person}{Ryan Julian}, \bibinfo{person}{Dmitry Kalashnikov}, \bibinfo{person}{Yuheng Kuang}, \bibinfo{person}{Isabel Leal}, \bibinfo{person}{Lisa Lee}, \bibinfo{person}{Tsang-Wei~Edward Lee}, \bibinfo{person}{Sergey Levine}, \bibinfo{person}{Yao Lu}, \bibinfo{person}{Henryk Michalewski},
  \bibinfo{person}{Igor Mordatch}, \bibinfo{person}{Karl Pertsch}, \bibinfo{person}{Kanishka Rao}, \bibinfo{person}{Krista Reymann}, \bibinfo{person}{Michael Ryoo}, \bibinfo{person}{Grecia Salazar}, \bibinfo{person}{Pannag Sanketi}, \bibinfo{person}{Pierre Sermanet}, \bibinfo{person}{Jaspiar Singh}, \bibinfo{person}{Anikait Singh}, \bibinfo{person}{Radu Soricut}, \bibinfo{person}{Huong Tran}, \bibinfo{person}{Vincent Vanhoucke}, \bibinfo{person}{Quan Vuong}, \bibinfo{person}{Ayzaan Wahid}, \bibinfo{person}{Stefan Welker}, \bibinfo{person}{Paul Wohlhart}, \bibinfo{person}{Jialin Wu}, \bibinfo{person}{Fei Xia}, \bibinfo{person}{Ted Xiao}, \bibinfo{person}{Peng Xu}, \bibinfo{person}{Sichun Xu}, \bibinfo{person}{Tianhe Yu}, {and} \bibinfo{person}{Brianna Zitkovich}.} \bibinfo{year}{2023}\natexlab{}.
\newblock \bibinfo{title}{RT-2: Vision-Language-Action Models Transfer Web Knowledge to Robotic Control}.
\newblock
\newblock
\showeprint[arxiv]{2307.15818}~[cs.RO]


\bibitem[Brown et~al\mbox{.}(2020)]%
        {ICL}
\bibfield{author}{\bibinfo{person}{Tom~B. Brown}, \bibinfo{person}{Benjamin Mann}, \bibinfo{person}{Nick Ryder}, \bibinfo{person}{Melanie Subbiah}, \bibinfo{person}{Jared Kaplan}, \bibinfo{person}{Prafulla Dhariwal}, \bibinfo{person}{Arvind Neelakantan}, \bibinfo{person}{Pranav Shyam}, \bibinfo{person}{Girish Sastry}, \bibinfo{person}{Amanda Askell}, \bibinfo{person}{Sandhini Agarwal}, \bibinfo{person}{Ariel Herbert-Voss}, \bibinfo{person}{Gretchen Krueger}, \bibinfo{person}{Tom Henighan}, \bibinfo{person}{Rewon Child}, \bibinfo{person}{Aditya Ramesh}, \bibinfo{person}{Daniel~M. Ziegler}, \bibinfo{person}{Jeffrey Wu}, \bibinfo{person}{Clemens Winter}, \bibinfo{person}{Christopher Hesse}, \bibinfo{person}{Mark Chen}, \bibinfo{person}{Eric Sigler}, \bibinfo{person}{Mateusz Litwin}, \bibinfo{person}{Scott Gray}, \bibinfo{person}{Benjamin Chess}, \bibinfo{person}{Jack Clark}, \bibinfo{person}{Christopher Berner}, \bibinfo{person}{Sam McCandlish}, \bibinfo{person}{Alec Radford}, \bibinfo{person}{Ilya Sutskever},
  {and} \bibinfo{person}{Dario Amodei}.} \bibinfo{year}{2020}\natexlab{}.
\newblock \bibinfo{title}{Language Models are Few-Shot Learners}.
\newblock
\newblock
\showeprint[arxiv]{2005.14165}~[cs.CL]


\bibitem[Chen et~al\mbox{.}(2023a)]%
        {driving_with_LLM}
\bibfield{author}{\bibinfo{person}{Long Chen}, \bibinfo{person}{Oleg Sinavski}, \bibinfo{person}{Jan Hünermann}, \bibinfo{person}{Alice Karnsund}, \bibinfo{person}{Andrew~James Willmott}, \bibinfo{person}{Danny Birch}, \bibinfo{person}{Daniel Maund}, {and} \bibinfo{person}{Jamie Shotton}.} \bibinfo{year}{2023}\natexlab{a}.
\newblock \bibinfo{title}{Driving with LLMs: Fusing Object-Level Vector Modality for Explainable Autonomous Driving}.
\newblock
\newblock
\showeprint[arxiv]{2310.01957}~[cs.RO]


\bibitem[Chen et~al\mbox{.}(2021)]%
        {chen2021evaluating}
\bibfield{author}{\bibinfo{person}{Mark Chen}, \bibinfo{person}{Jerry Tworek}, \bibinfo{person}{Heewoo Jun}, \bibinfo{person}{Qiming Yuan}, \bibinfo{person}{Henrique~Ponde de Oliveira~Pinto}, \bibinfo{person}{Jared Kaplan}, \bibinfo{person}{Harri Edwards}, \bibinfo{person}{Yuri Burda}, \bibinfo{person}{Nicholas Joseph}, \bibinfo{person}{Greg Brockman}, \bibinfo{person}{Alex Ray}, \bibinfo{person}{Raul Puri}, \bibinfo{person}{Gretchen Krueger}, \bibinfo{person}{Michael Petrov}, \bibinfo{person}{Heidy Khlaaf}, \bibinfo{person}{Girish Sastry}, \bibinfo{person}{Pamela Mishkin}, \bibinfo{person}{Brooke Chan}, \bibinfo{person}{Scott Gray}, \bibinfo{person}{Nick Ryder}, \bibinfo{person}{Mikhail Pavlov}, \bibinfo{person}{Alethea Power}, \bibinfo{person}{Lukasz Kaiser}, \bibinfo{person}{Mohammad Bavarian}, \bibinfo{person}{Clemens Winter}, \bibinfo{person}{Philippe Tillet}, \bibinfo{person}{Felipe~Petroski Such}, \bibinfo{person}{Dave Cummings}, \bibinfo{person}{Matthias Plappert}, \bibinfo{person}{Fotios Chantzis},
  \bibinfo{person}{Elizabeth Barnes}, \bibinfo{person}{Ariel Herbert-Voss}, \bibinfo{person}{William~Hebgen Guss}, \bibinfo{person}{Alex Nichol}, \bibinfo{person}{Alex Paino}, \bibinfo{person}{Nikolas Tezak}, \bibinfo{person}{Jie Tang}, \bibinfo{person}{Igor Babuschkin}, \bibinfo{person}{Suchir Balaji}, \bibinfo{person}{Shantanu Jain}, \bibinfo{person}{William Saunders}, \bibinfo{person}{Christopher Hesse}, \bibinfo{person}{Andrew~N. Carr}, \bibinfo{person}{Jan Leike}, \bibinfo{person}{Josh Achiam}, \bibinfo{person}{Vedant Misra}, \bibinfo{person}{Evan Morikawa}, \bibinfo{person}{Alec Radford}, \bibinfo{person}{Matthew Knight}, \bibinfo{person}{Miles Brundage}, \bibinfo{person}{Mira Murati}, \bibinfo{person}{Katie Mayer}, \bibinfo{person}{Peter Welinder}, \bibinfo{person}{Bob McGrew}, \bibinfo{person}{Dario Amodei}, \bibinfo{person}{Sam McCandlish}, \bibinfo{person}{Ilya Sutskever}, {and} \bibinfo{person}{Wojciech Zaremba}.} \bibinfo{year}{2021}\natexlab{}.
\newblock \bibinfo{title}{Evaluating Large Language Models Trained on Code}.
\newblock
\newblock
\showeprint[arxiv]{2107.03374}~[cs.LG]


\bibitem[Chen et~al\mbox{.}(2023b)]%
        {chen2023agentverse}
\bibfield{author}{\bibinfo{person}{Weize Chen}, \bibinfo{person}{Yusheng Su}, \bibinfo{person}{Jingwei Zuo}, \bibinfo{person}{Cheng Yang}, \bibinfo{person}{Chenfei Yuan}, \bibinfo{person}{Chi-Min Chan}, \bibinfo{person}{Heyang Yu}, \bibinfo{person}{Yaxi Lu}, \bibinfo{person}{Yi-Hsin Hung}, \bibinfo{person}{Chen Qian}, \bibinfo{person}{Yujia Qin}, \bibinfo{person}{Xin Cong}, \bibinfo{person}{Ruobing Xie}, \bibinfo{person}{Zhiyuan Liu}, \bibinfo{person}{Maosong Sun}, {and} \bibinfo{person}{Jie Zhou}.} \bibinfo{year}{2023}\natexlab{b}.
\newblock \bibinfo{title}{AgentVerse: Facilitating Multi-Agent Collaboration and Exploring Emergent Behaviors}.
\newblock
\newblock
\showeprint[arxiv]{2308.10848}~[cs.CL]


\bibitem[Cheng et~al\mbox{.}(2009)]%
        {Cheng2009}
\bibfield{author}{\bibinfo{person}{Betty H.~C. Cheng} {et~al\mbox{.}}} \bibinfo{year}{2009}\natexlab{}.
\newblock \bibinfo{booktitle}{\emph{Software Engineering for Self-Adaptive Systems: A Research Roadmap}}.
\newblock \bibinfo{publisher}{Springer}, \bibinfo{pages}{1--26}.
\newblock


\bibitem[Cleland-Huang et~al\mbox{.}(2022)]%
        {MAPEK_HMTeaming}
\bibfield{author}{\bibinfo{person}{Jane Cleland-Huang}, \bibinfo{person}{Ankit Agrawal}, \bibinfo{person}{Michael Vierhauser}, \bibinfo{person}{Michael Murphy}, {and} \bibinfo{person}{Mike Prieto}.} \bibinfo{year}{2022}\natexlab{}.
\newblock \showarticletitle{Extending MAPE-K to Support Human-Machine Teaming}. In \bibinfo{booktitle}{\emph{Proceedings of the 17th Symposium on Software Engineering for Adaptive and Self-Managing Systems}} (Pittsburgh, Pennsylvania) \emph{(\bibinfo{series}{SEAMS '22})}. \bibinfo{publisher}{Association for Computing Machinery}, \bibinfo{address}{New York, NY, USA}, \bibinfo{pages}{120–131}.
\newblock
\showISBNx{9781450393058}


\bibitem[Cosler et~al\mbox{.}(2023)]%
        {LLM_req_translation}
\bibfield{author}{\bibinfo{person}{Matthias Cosler}, \bibinfo{person}{Christopher Hahn}, \bibinfo{person}{Daniel Mendoza}, \bibinfo{person}{Frederik Schmitt}, {and} \bibinfo{person}{Caroline Trippel}.} \bibinfo{year}{2023}\natexlab{}.
\newblock \showarticletitle{nl2spec: Interactively Translating Unstructured Natural Language to Temporal Logics with Large Language Models}. In \bibinfo{booktitle}{\emph{Computer Aided Verification}}. \bibinfo{publisher}{Springer Nature Switzerland}, \bibinfo{pages}{383--396}.
\newblock
\showISBNx{978-3-031-37703-7}


\bibitem[de~Zarzà et~al\mbox{.}(2023)]%
        {LLM_adaptivePID}
\bibfield{author}{\bibinfo{person}{I. de Zarzà}, \bibinfo{person}{J. de Curtò}, \bibinfo{person}{Gemma Roig}, {and} \bibinfo{person}{Carlos~T. Calafate}.} \bibinfo{year}{2023}\natexlab{}.
\newblock \showarticletitle{LLM Adaptive PID Control for B5G Truck Platooning Systems}.
\newblock \bibinfo{journal}{\emph{Sensors}} \bibinfo{volume}{23}, \bibinfo{number}{13} (\bibinfo{year}{2023}).
\newblock
\showISSN{1424-8220}


\bibitem[Devlin et~al\mbox{.}(2019)]%
        {devlin2019bert}
\bibfield{author}{\bibinfo{person}{Jacob Devlin}, \bibinfo{person}{Ming-Wei Chang}, \bibinfo{person}{Kenton Lee}, {and} \bibinfo{person}{Kristina Toutanova}.} \bibinfo{year}{2019}\natexlab{}.
\newblock \bibinfo{title}{BERT: Pre-training of Deep Bidirectional Transformers for Language Understanding}.
\newblock , \bibinfo{numpages}{4171--4186}~pages.
\newblock


\bibitem[Ding et~al\mbox{.}(2023)]%
        {ding2023hilmd}
\bibfield{author}{\bibinfo{person}{Xinpeng Ding}, \bibinfo{person}{Jianhua Han}, \bibinfo{person}{Hang Xu}, \bibinfo{person}{Wei Zhang}, {and} \bibinfo{person}{Xiaomeng Li}.} \bibinfo{year}{2023}\natexlab{}.
\newblock \bibinfo{title}{HiLM-D: Towards High-Resolution Understanding in Multimodal Large Language Models for Autonomous Driving}.
\newblock
\newblock
\showeprint[arxiv]{2309.05186}~[cs.CV]


\bibitem[Du et~al\mbox{.}(2023)]%
        {du2023improving}
\bibfield{author}{\bibinfo{person}{Yilun Du}, \bibinfo{person}{Shuang Li}, \bibinfo{person}{Antonio Torralba}, \bibinfo{person}{Joshua~B. Tenenbaum}, {and} \bibinfo{person}{Igor Mordatch}.} \bibinfo{year}{2023}\natexlab{}.
\newblock \bibinfo{title}{Improving Factuality and Reasoning in Language Models through Multiagent Debate}.
\newblock
\newblock
\showeprint[arxiv]{2305.14325}~[cs.CL]


\bibitem[Fan et~al\mbox{.}(2023)]%
        {LLM4SE_survey2}
\bibfield{author}{\bibinfo{person}{Angela Fan}, \bibinfo{person}{Beliz Gokkaya}, \bibinfo{person}{Mark Harman}, \bibinfo{person}{Mitya Lyubarskiy}, \bibinfo{person}{Shubho Sengupta}, \bibinfo{person}{Shin Yoo}, {and} \bibinfo{person}{Jie~M. Zhang}.} \bibinfo{year}{2023}\natexlab{}.
\newblock \bibinfo{title}{Large Language Models for Software Engineering: Survey and Open Problems}.
\newblock
\newblock
\showeprint[arxiv]{2310.03533}~[cs.SE]


\bibitem[Garlan et~al\mbox{.}(2004)]%
        {1350726}
\bibfield{author}{\bibinfo{person}{D. Garlan} {et~al\mbox{.}}} \bibinfo{year}{2004}\natexlab{}.
\newblock \showarticletitle{Rainbow: architecture-based self-adaptation with reusable infrastructure}.
\newblock \bibinfo{journal}{\emph{Computer}} \bibinfo{volume}{37}, \bibinfo{number}{10} (\bibinfo{year}{2004}).
\newblock


\bibitem[Gong et~al\mbox{.}(2023)]%
        {gong2023mindagent}
\bibfield{author}{\bibinfo{person}{Ran Gong}, \bibinfo{person}{Qiuyuan Huang}, \bibinfo{person}{Xiaojian Ma}, \bibinfo{person}{Hoi Vo}, \bibinfo{person}{Zane Durante}, \bibinfo{person}{Yusuke Noda}, \bibinfo{person}{Zilong Zheng}, \bibinfo{person}{Song-Chun Zhu}, \bibinfo{person}{Demetri Terzopoulos}, \bibinfo{person}{Li Fei-Fei}, {and} \bibinfo{person}{Jianfeng Gao}.} \bibinfo{year}{2023}\natexlab{}.
\newblock \bibinfo{title}{MindAgent: Emergent Gaming Interaction}.
\newblock
\newblock
\showeprint[arxiv]{2309.09971}~[cs.AI]


\bibitem[Graule and Isler(2023)]%
        {graule2023ggllm}
\bibfield{author}{\bibinfo{person}{Moritz~A. Graule} {and} \bibinfo{person}{Volkan Isler}.} \bibinfo{year}{2023}\natexlab{}.
\newblock \bibinfo{title}{GG-LLM: Geometrically Grounding Large Language Models for Zero-shot Human Activity Forecasting in Human-Aware Task Planning}.
\newblock
\newblock
\showeprint[arxiv]{2310.20034}~[cs.RO]


\bibitem[Hao et~al\mbox{.}(2023)]%
        {hao2023reasoning}
\bibfield{author}{\bibinfo{person}{Shibo Hao}, \bibinfo{person}{Yi Gu}, \bibinfo{person}{Haodi Ma}, \bibinfo{person}{Joshua~Jiahua Hong}, \bibinfo{person}{Zhen Wang}, \bibinfo{person}{Daisy~Zhe Wang}, {and} \bibinfo{person}{Zhiting Hu}.} \bibinfo{year}{2023}\natexlab{}.
\newblock \bibinfo{title}{Reasoning with Language Model is Planning with World Model}.
\newblock
\newblock
\showeprint[arxiv]{2305.14992}~[cs.CL]


\bibitem[Hiroyuki~Nakagawa(2023)]%
        {LLM_goalmodel}
\bibfield{author}{\bibinfo{person}{Shinichi~Honiden Hiroyuki~Nakagawa}.} \bibinfo{year}{2023}\natexlab{}.
\newblock \showarticletitle{MAPE-K Loop-based Goal Model Generation Using Generative AI}. In \bibinfo{booktitle}{\emph{IEEE 31st International Requirements Engineering Conference Workshop}}.
\newblock


\bibitem[Hollmann et~al\mbox{.}(2023)]%
        {LLM_feature_enginering}
\bibfield{author}{\bibinfo{person}{Noah Hollmann}, \bibinfo{person}{Samuel Müller}, {and} \bibinfo{person}{Frank Hutter}.} \bibinfo{year}{2023}\natexlab{}.
\newblock \bibinfo{title}{Large Language Models for Automated Data Science: Introducing CAAFE for Context-Aware Automated Feature Engineering}.
\newblock
\newblock
\showeprint[arxiv]{2305.03403}~[cs.AI]


\bibitem[Hong et~al\mbox{.}(2023)]%
        {metaGPT}
\bibfield{author}{\bibinfo{person}{Sirui Hong}, \bibinfo{person}{Xiawu Zheng}, \bibinfo{person}{Jonathan Chen}, \bibinfo{person}{Yuheng Cheng}, \bibinfo{person}{Jinlin Wang}, \bibinfo{person}{Ceyao Zhang}, \bibinfo{person}{Zili Wang}, \bibinfo{person}{Steven Ka~Shing Yau}, \bibinfo{person}{Zijuan Lin}, \bibinfo{person}{Liyang Zhou}, \bibinfo{person}{Chenyu Ran}, \bibinfo{person}{Lingfeng Xiao}, {and} \bibinfo{person}{Chenglin Wu}.} \bibinfo{year}{2023}\natexlab{}.
\newblock \bibinfo{title}{MetaGPT: Meta Programming for Multi-Agent Collaborative Framework}.
\newblock
\newblock
\showeprint[arxiv]{2308.00352}~[cs.AI]


\bibitem[Hou et~al\mbox{.}(2023)]%
        {LLMonSE_Survey1}
\bibfield{author}{\bibinfo{person}{Xinyi Hou}, \bibinfo{person}{Yanjie Zhao}, \bibinfo{person}{Yue Liu}, \bibinfo{person}{Zhou Yang}, \bibinfo{person}{Kailong Wang}, \bibinfo{person}{Li Li}, \bibinfo{person}{Xiapu Luo}, \bibinfo{person}{David Lo}, \bibinfo{person}{John Grundy}, {and} \bibinfo{person}{Haoyu Wang}.} \bibinfo{year}{2023}\natexlab{}.
\newblock \bibinfo{title}{Large Language Models for Software Engineering: A Systematic Literature Review}.
\newblock
\newblock
\showeprint[arxiv]{2308.10620}~[cs.SE]


\bibitem[Howard and Ruder(2018)]%
        {LMFineTuning}
\bibfield{author}{\bibinfo{person}{Jeremy Howard} {and} \bibinfo{person}{Sebastian Ruder}.} \bibinfo{year}{2018}\natexlab{}.
\newblock \showarticletitle{Universal Language Model Fine-tuning for Text Classification}. In \bibinfo{booktitle}{\emph{Proceedings of the 56th Annual Meeting of the Association for Computational Linguistics (Volume 1: Long Papers)}}, \bibfield{editor}{\bibinfo{person}{Iryna Gurevych} {and} \bibinfo{person}{Yusuke Miyao}} (Eds.). \bibinfo{publisher}{Association for Computational Linguistics}, \bibinfo{address}{Melbourne, Australia}, \bibinfo{pages}{328--339}.
\newblock


\bibitem[Hu et~al\mbox{.}(2023)]%
        {GATA1}
\bibfield{author}{\bibinfo{person}{Anthony Hu}, \bibinfo{person}{Lloyd Russell}, \bibinfo{person}{Hudson Yeo}, \bibinfo{person}{Zak Murez}, \bibinfo{person}{George Fedoseev}, \bibinfo{person}{Alex Kendall}, \bibinfo{person}{Jamie Shotton}, {and} \bibinfo{person}{Gianluca Corrado}.} \bibinfo{year}{2023}\natexlab{}.
\newblock \bibinfo{title}{GAIA-1: A Generative World Model for Autonomous Driving}.
\newblock
\newblock
\showeprint[arxiv]{2309.17080}~[cs.CV]


\bibitem[Kannan et~al\mbox{.}(2023)]%
        {kannan2023smartllm}
\bibfield{author}{\bibinfo{person}{Shyam~Sundar Kannan}, \bibinfo{person}{Vishnunandan L.~N. Venkatesh}, {and} \bibinfo{person}{Byung-Cheol Min}.} \bibinfo{year}{2023}\natexlab{}.
\newblock \bibinfo{title}{SMART-LLM: Smart Multi-Agent Robot Task Planning using Large Language Models}.
\newblock
\newblock
\showeprint[arxiv]{2309.10062}~[cs.RO]


\bibitem[Kaplan et~al\mbox{.}(2020)]%
        {kaplan2020scaling}
\bibfield{author}{\bibinfo{person}{Jared Kaplan}, \bibinfo{person}{Sam McCandlish}, \bibinfo{person}{Tom Henighan}, \bibinfo{person}{Tom~B. Brown}, \bibinfo{person}{Benjamin Chess}, \bibinfo{person}{Rewon Child}, \bibinfo{person}{Scott Gray}, \bibinfo{person}{Alec Radford}, \bibinfo{person}{Jeffrey Wu}, {and} \bibinfo{person}{Dario Amodei}.} \bibinfo{year}{2020}\natexlab{}.
\newblock \bibinfo{title}{Scaling Laws for Neural Language Models}.
\newblock
\newblock
\showeprint[arxiv]{2001.08361}~[cs.LG]


\bibitem[{Kephart} and David{Chess}(2003)]%
        {MAPE}
\bibfield{author}{\bibinfo{person}{Jeff {Kephart}} {and} \bibinfo{person}{David{Chess}}.} \bibinfo{year}{2003}\natexlab{}.
\newblock \showarticletitle{The vision of autonomic computing}.
\newblock \bibinfo{journal}{\emph{Computer}} \bibinfo{volume}{36}, \bibinfo{number}{1} (\bibinfo{date}{Jan} \bibinfo{year}{2003}), \bibinfo{pages}{41--50}.
\newblock


\bibitem[Li et~al\mbox{.}(2020)]%
        {nianyu_seams20}
\bibfield{author}{\bibinfo{person}{Nianyu Li}, \bibinfo{person}{Sridhar Adepu}, \bibinfo{person}{Eunsuk Kang}, {and} \bibinfo{person}{David Garlan}.} \bibinfo{year}{2020}\natexlab{}.
\newblock \showarticletitle{Explanations for Human-on-the-Loop: A Probabilistic Model Checking Approach}. In \bibinfo{booktitle}{\emph{Proceedings of the IEEE/ACM 15th International Symposium on Software Engineering for Adaptive and Self-Managing Systems}} \emph{(\bibinfo{series}{SEAMS '20})}. \bibinfo{pages}{181–187}.
\newblock


\bibitem[Li et~al\mbox{.}(2023)]%
        {seams23_preference}
\bibfield{author}{\bibinfo{person}{Nianyu Li}, \bibinfo{person}{Mingyue Zhang}, \bibinfo{person}{Jialong Li}, \bibinfo{person}{Eunsuk Kang}, {and} \bibinfo{person}{Kenji Tei}.} \bibinfo{year}{2023}\natexlab{}.
\newblock \showarticletitle{Preference Adaptation: user satisfaction is all you need!}. In \bibinfo{booktitle}{\emph{2023 IEEE/ACM 18th Symposium on Software Engineering for Adaptive and Self-Managing Systems (SEAMS)}}. \bibinfo{pages}{133--144}.
\newblock


\bibitem[Liu et~al\mbox{.}(2023)]%
        {liu2023chipnemo}
\bibfield{author}{\bibinfo{person}{Mingjie Liu}, \bibinfo{person}{Teo Ene}, \bibinfo{person}{Robert Kirby}, \bibinfo{person}{Chris Cheng}, \bibinfo{person}{Nathaniel Pinckney}, \bibinfo{person}{Rongjian Liang}, \bibinfo{person}{Jonah Alben}, \bibinfo{person}{Himyanshu Anand}, \bibinfo{person}{Sanmitra Banerjee}, \bibinfo{person}{Ismet Bayraktaroglu}, \bibinfo{person}{Bonita Bhaskaran}, \bibinfo{person}{Bryan Catanzaro}, \bibinfo{person}{Arjun Chaudhuri}, \bibinfo{person}{Sharon Clay}, \bibinfo{person}{Bill Dally}, \bibinfo{person}{Laura Dang}, \bibinfo{person}{Parikshit Deshpande}, \bibinfo{person}{Siddhanth Dhodhi}, \bibinfo{person}{Sameer Halepete}, \bibinfo{person}{Eric Hill}, \bibinfo{person}{Jiashang Hu}, \bibinfo{person}{Sumit Jain}, \bibinfo{person}{Brucek Khailany}, \bibinfo{person}{Kishor Kunal}, \bibinfo{person}{Xiaowei Li}, \bibinfo{person}{Hao Liu}, \bibinfo{person}{Stuart Oberman}, \bibinfo{person}{Sujeet Omar}, \bibinfo{person}{Sreedhar Pratty}, \bibinfo{person}{Ambar Sarkar},
  \bibinfo{person}{Zhengjiang Shao}, \bibinfo{person}{Hanfei Sun}, \bibinfo{person}{Pratik~P Suthar}, \bibinfo{person}{Varun Tej}, \bibinfo{person}{Kaizhe Xu}, {and} \bibinfo{person}{Haoxing Ren}.} \bibinfo{year}{2023}\natexlab{}.
\newblock \bibinfo{title}{ChipNeMo: Domain-Adapted LLMs for Chip Design}.
\newblock
\newblock
\showeprint[arxiv]{2311.00176}~[cs.CL]


\bibitem[Lyu et~al\mbox{.}(2023)]%
        {lyu2023macawllm}
\bibfield{author}{\bibinfo{person}{Chenyang Lyu}, \bibinfo{person}{Minghao Wu}, \bibinfo{person}{Longyue Wang}, \bibinfo{person}{Xinting Huang}, \bibinfo{person}{Bingshuai Liu}, \bibinfo{person}{Zefeng Du}, \bibinfo{person}{Shuming Shi}, {and} \bibinfo{person}{Zhaopeng Tu}.} \bibinfo{year}{2023}\natexlab{}.
\newblock \bibinfo{title}{Macaw-LLM: Multi-Modal Language Modeling with Image, Audio, Video, and Text Integration}.
\newblock
\newblock
\showeprint[arxiv]{2306.09093}~[cs.CL]


\bibitem[Ma et~al\mbox{.}(2023)]%
        {ma2023chatgpt}
\bibfield{author}{\bibinfo{person}{Wei Ma}, \bibinfo{person}{Shangqing Liu}, \bibinfo{person}{Wenhan Wang}, \bibinfo{person}{Qiang Hu}, \bibinfo{person}{Ye Liu}, \bibinfo{person}{Cen Zhang}, \bibinfo{person}{Liming Nie}, {and} \bibinfo{person}{Yang Liu}.} \bibinfo{year}{2023}\natexlab{}.
\newblock \bibinfo{title}{ChatGPT: Understanding Code Syntax and Semantics}.
\newblock
\newblock
\showeprint[arxiv]{2305.12138}~[cs.SE]


\bibitem[Maddigan and Susnjak(2023)]%
        {Chat2VIS}
\bibfield{author}{\bibinfo{person}{Paula Maddigan} {and} \bibinfo{person}{Teo Susnjak}.} \bibinfo{year}{2023}\natexlab{}.
\newblock \showarticletitle{Chat2VIS: Generating Data Visualizations via Natural Language Using ChatGPT, Codex and GPT-3 Large Language Models}.
\newblock \bibinfo{journal}{\emph{IEEE Access}}  \bibinfo{volume}{11} (\bibinfo{year}{2023}).
\newblock


\bibitem[Mandal et~al\mbox{.}(2023)]%
        {LLM_specification_synthesis}
\bibfield{author}{\bibinfo{person}{Shantanu Mandal}, \bibinfo{person}{Adhrik Chethan}, \bibinfo{person}{Vahid Janfaza}, \bibinfo{person}{S~M~Farabi Mahmud}, \bibinfo{person}{Todd~A Anderson}, \bibinfo{person}{Javier Turek}, \bibinfo{person}{Jesmin~Jahan Tithi}, {and} \bibinfo{person}{Abdullah Muzahid}.} \bibinfo{year}{2023}\natexlab{}.
\newblock \bibinfo{title}{Large Language Models Based Automatic Synthesis of Software Specifications}.
\newblock
\newblock
\showeprint[arxiv]{2304.09181}~[cs.SE]


\bibitem[Mao et~al\mbox{.}(2023)]%
        {mao2023alympics}
\bibfield{author}{\bibinfo{person}{Shaoguang Mao}, \bibinfo{person}{Yuzhe Cai}, \bibinfo{person}{Yan Xia}, \bibinfo{person}{Wenshan Wu}, \bibinfo{person}{Xun Wang}, \bibinfo{person}{Fengyi Wang}, \bibinfo{person}{Tao Ge}, {and} \bibinfo{person}{Furu Wei}.} \bibinfo{year}{2023}\natexlab{}.
\newblock \bibinfo{title}{ALYMPICS: Language Agents Meet Game Theory}.
\newblock
\newblock
\showeprint[arxiv]{2311.03220}~[cs.CL]


\bibitem[Mingyue et~al\mbox{.}(2020)]%
        {zmy_xuebao20}
\bibfield{author}{\bibinfo{person}{Zhang Mingyue}, \bibinfo{person}{Jin Zhi}, \bibinfo{person}{Zhao Haiyan}, {and} \bibinfo{person}{Luo Yixing}.} \bibinfo{year}{2020}\natexlab{}.
\newblock \showarticletitle{Survey of Machine Learning Enabled Software Self-adaptation}.
\newblock \bibinfo{journal}{\emph{Journal of Software (in Chinese)}} \bibinfo{volume}{31}, \bibinfo{number}{8} (\bibinfo{year}{2020}), \bibinfo{pages}{2404--2431}.
\newblock


\bibitem[Nourbakhsh and Bang(2019)]%
        {LM_finance}
\bibfield{author}{\bibinfo{person}{Armineh Nourbakhsh} {and} \bibinfo{person}{Grace Bang}.} \bibinfo{year}{2019}\natexlab{}.
\newblock \bibinfo{title}{A framework for anomaly detection using language modeling, and its applications to finance}.
\newblock
\newblock
\showeprint[arxiv]{1908.09156}~[cs.CL]


\bibitem[OpenAI(2023)]%
        {openai2023gpt4}
\bibfield{author}{\bibinfo{person}{OpenAI}.} \bibinfo{year}{2023}\natexlab{}.
\newblock \bibinfo{title}{GPT-4 Technical Report}.
\newblock
\newblock
\showeprint[arxiv]{2303.08774}~[cs.CL]


\bibitem[Qin et~al\mbox{.}(2023a)]%
        {qin2023tool}
\bibfield{author}{\bibinfo{person}{Yujia Qin}, \bibinfo{person}{Shengding Hu}, \bibinfo{person}{Yankai Lin}, \bibinfo{person}{Weize Chen}, \bibinfo{person}{Ning Ding}, \bibinfo{person}{Ganqu Cui}, \bibinfo{person}{Zheni Zeng}, \bibinfo{person}{Yufei Huang}, \bibinfo{person}{Chaojun Xiao}, \bibinfo{person}{Chi Han}, \bibinfo{person}{Yi~Ren Fung}, \bibinfo{person}{Yusheng Su}, \bibinfo{person}{Huadong Wang}, \bibinfo{person}{Cheng Qian}, \bibinfo{person}{Runchu Tian}, \bibinfo{person}{Kunlun Zhu}, \bibinfo{person}{Shihao Liang}, \bibinfo{person}{Xingyu Shen}, \bibinfo{person}{Bokai Xu}, \bibinfo{person}{Zhen Zhang}, \bibinfo{person}{Yining Ye}, \bibinfo{person}{Bowen Li}, \bibinfo{person}{Ziwei Tang}, \bibinfo{person}{Jing Yi}, \bibinfo{person}{Yuzhang Zhu}, \bibinfo{person}{Zhenning Dai}, \bibinfo{person}{Lan Yan}, \bibinfo{person}{Xin Cong}, \bibinfo{person}{Yaxi Lu}, \bibinfo{person}{Weilin Zhao}, \bibinfo{person}{Yuxiang Huang}, \bibinfo{person}{Junxi Yan}, \bibinfo{person}{Xu Han}, \bibinfo{person}{Xian
  Sun}, \bibinfo{person}{Dahai Li}, \bibinfo{person}{Jason Phang}, \bibinfo{person}{Cheng Yang}, \bibinfo{person}{Tongshuang Wu}, \bibinfo{person}{Heng Ji}, \bibinfo{person}{Zhiyuan Liu}, {and} \bibinfo{person}{Maosong Sun}.} \bibinfo{year}{2023}\natexlab{a}.
\newblock \bibinfo{title}{Tool Learning with Foundation Models}.
\newblock
\newblock
\showeprint[arxiv]{2304.08354}~[cs.CL]


\bibitem[Qin et~al\mbox{.}(2023b)]%
        {qin2023toolllm}
\bibfield{author}{\bibinfo{person}{Yujia Qin}, \bibinfo{person}{Shihao Liang}, \bibinfo{person}{Yining Ye}, \bibinfo{person}{Kunlun Zhu}, \bibinfo{person}{Lan Yan}, \bibinfo{person}{Yaxi Lu}, \bibinfo{person}{Yankai Lin}, \bibinfo{person}{Xin Cong}, \bibinfo{person}{Xiangru Tang}, \bibinfo{person}{Bill Qian}, \bibinfo{person}{Sihan Zhao}, \bibinfo{person}{Lauren Hong}, \bibinfo{person}{Runchu Tian}, \bibinfo{person}{Ruobing Xie}, \bibinfo{person}{Jie Zhou}, \bibinfo{person}{Mark Gerstein}, \bibinfo{person}{Dahai Li}, \bibinfo{person}{Zhiyuan Liu}, {and} \bibinfo{person}{Maosong Sun}.} \bibinfo{year}{2023}\natexlab{b}.
\newblock \bibinfo{title}{ToolLLM: Facilitating Large Language Models to Master 16000+ Real-world APIs}.
\newblock
\newblock
\showeprint[arxiv]{2307.16789}~[cs.AI]


\bibitem[Rana et~al\mbox{.}(2023)]%
        {SayPlan}
\bibfield{author}{\bibinfo{person}{Krishan Rana}, \bibinfo{person}{Jesse Haviland}, \bibinfo{person}{Sourav Garg}, \bibinfo{person}{Jad Abou-Chakra}, \bibinfo{person}{Ian Reid}, {and} \bibinfo{person}{Niko Suenderhauf}.} \bibinfo{year}{2023}\natexlab{}.
\newblock \bibinfo{title}{SayPlan: Grounding Large Language Models using 3D Scene Graphs for Scalable Robot Task Planning}.
\newblock
\newblock
\showeprint[arxiv]{2307.06135}~[cs.RO]


\bibitem[Schirner et~al\mbox{.}(2013)]%
        {HITL_CPS}
\bibfield{author}{\bibinfo{person}{Gunar Schirner}, \bibinfo{person}{Deniz Erdogmus}, \bibinfo{person}{Kaushik Chowdhury}, {and} \bibinfo{person}{Taskin Padir}.} \bibinfo{year}{2013}\natexlab{}.
\newblock \showarticletitle{The Future of Human-in-the-Loop Cyber-Physical Systems}.
\newblock \bibinfo{journal}{\emph{Computer}} \bibinfo{volume}{46}, \bibinfo{number}{1} (\bibinfo{year}{2013}), \bibinfo{pages}{36--45}.
\newblock


\bibitem[Seff et~al\mbox{.}(2023)]%
        {seff2023motionlm}
\bibfield{author}{\bibinfo{person}{Ari Seff}, \bibinfo{person}{Brian Cera}, \bibinfo{person}{Dian Chen}, \bibinfo{person}{Mason Ng}, \bibinfo{person}{Aurick Zhou}, \bibinfo{person}{Nigamaa Nayakanti}, \bibinfo{person}{Khaled~S. Refaat}, \bibinfo{person}{Rami Al-Rfou}, {and} \bibinfo{person}{Benjamin Sapp}.} \bibinfo{year}{2023}\natexlab{}.
\newblock \bibinfo{title}{MotionLM: Multi-Agent Motion Forecasting as Language Modeling}.
\newblock
\newblock
\showeprint[arxiv]{2309.16534}~[cs.CV]


\bibitem[Song et~al\mbox{.}(2023)]%
        {song2023selfrefined}
\bibfield{author}{\bibinfo{person}{Jiayang Song}, \bibinfo{person}{Zhehua Zhou}, \bibinfo{person}{Jiawei Liu}, \bibinfo{person}{Chunrong Fang}, \bibinfo{person}{Zhan Shu}, {and} \bibinfo{person}{Lei Ma}.} \bibinfo{year}{2023}\natexlab{}.
\newblock \bibinfo{title}{Self-Refined Large Language Model as Automated Reward Function Designer for Deep Reinforcement Learning in Robotics}.
\newblock
\newblock
\showeprint[arxiv]{2309.06687}~[cs.RO]


\bibitem[Stewart et~al\mbox{.}(2023)]%
        {stewart2023FMC}
\bibfield{author}{\bibinfo{person}{Michael Stewart}, \bibinfo{person}{Melinda Hodkiewicz}, {and} \bibinfo{person}{Sirui Li}.} \bibinfo{year}{2023}\natexlab{}.
\newblock \bibinfo{title}{Large Language Models for Failure Mode Classification: An Investigation}.
\newblock
\newblock
\showeprint[arxiv]{2309.08181}~[cs.CL]


\bibitem[Sykes et~al\mbox{.}(2008)]%
        {3_layer_model}
\bibfield{author}{\bibinfo{person}{Daniel Sykes}, \bibinfo{person}{William Heaven}, \bibinfo{person}{Jeff Magee}, {and} \bibinfo{person}{Jeff Kramer}.} \bibinfo{year}{2008}\natexlab{}.
\newblock \showarticletitle{From Goals to Components: A Combined Approach to Self-Management}. In \bibinfo{booktitle}{\emph{Proceedings of the 2008 International Workshop on Software Engineering for Adaptive and Self-Managing Systems}} \emph{(\bibinfo{series}{SEAMS '08})}. \bibinfo{pages}{1–8}.
\newblock
\showISBNx{9781605580371}


\bibitem[Tagliabue et~al\mbox{.}(2023)]%
        {tagliabue2023real}
\bibfield{author}{\bibinfo{person}{Andrea Tagliabue}, \bibinfo{person}{Kota Kondo}, \bibinfo{person}{Tong Zhao}, \bibinfo{person}{Mason Peterson}, \bibinfo{person}{Claudius~T. Tewari}, {and} \bibinfo{person}{Jonathan~P. How}.} \bibinfo{year}{2023}\natexlab{}.
\newblock \bibinfo{title}{REAL: Resilience and Adaptation using Large Language Models on Autonomous Aerial Robots}.
\newblock
\newblock
\showeprint[arxiv]{2311.01403}~[cs.RO]


\bibitem[Tang et~al\mbox{.}(2023)]%
        {LLM_AD_QA1}
\bibfield{author}{\bibinfo{person}{Yun Tang}, \bibinfo{person}{Antonio A.~Bruto da Costa}, \bibinfo{person}{Jason Zhang}, \bibinfo{person}{Irvine Patrick}, \bibinfo{person}{Siddartha Khastgir}, {and} \bibinfo{person}{Paul Jennings}.} \bibinfo{year}{2023}\natexlab{}.
\newblock \bibinfo{title}{Domain Knowledge Distillation from Large Language Model: An Empirical Study in the Autonomous Driving Domain}.
\newblock
\newblock
\showeprint[arxiv]{2307.11769}~[cs.CL]


\bibitem[Thomas et~al\mbox{.}(2023)]%
        {thomas2023large}
\bibfield{author}{\bibinfo{person}{Paul Thomas}, \bibinfo{person}{Seth Spielman}, \bibinfo{person}{Nick Craswell}, {and} \bibinfo{person}{Bhaskar Mitra}.} \bibinfo{year}{2023}\natexlab{}.
\newblock \bibinfo{title}{Large language models can accurately predict searcher preferences}.
\newblock
\newblock
\showeprint[arxiv]{2309.10621}~[cs.IR]


\bibitem[Tsigkanos et~al\mbox{.}(2023)]%
        {LLM_test_generation}
\bibfield{author}{\bibinfo{person}{Christos Tsigkanos}, \bibinfo{person}{Pooja Rani}, \bibinfo{person}{Sebastian M{\"u}ller}, {and} \bibinfo{person}{Timo Kehrer}.} \bibinfo{year}{2023}\natexlab{}.
\newblock \showarticletitle{Variable Discovery with Large Language Models for Metamorphic Testing of Scientific Software}. In \bibinfo{booktitle}{\emph{Computational Science -- ICCS 2023}}. \bibinfo{publisher}{Springer Nature Switzerland}, \bibinfo{address}{Cham}, \bibinfo{pages}{321--335}.
\newblock
\showISBNx{978-3-031-35995-8}


\bibitem[Tufano et~al\mbox{.}(2022)]%
        {LLM_code_review}
\bibfield{author}{\bibinfo{person}{Rosalia Tufano}, \bibinfo{person}{Simone Masiero}, \bibinfo{person}{Antonio Mastropaolo}, \bibinfo{person}{Luca Pascarella}, \bibinfo{person}{Denys Poshyvanyk}, {and} \bibinfo{person}{Gabriele Bavota}.} \bibinfo{year}{2022}\natexlab{}.
\newblock \showarticletitle{Using Pre-Trained Models to Boost Code Review Automation}. In \bibinfo{booktitle}{\emph{Proceedings of the 44th International Conference on Software Engineering}} (Pittsburgh, Pennsylvania) \emph{(\bibinfo{series}{ICSE '22})}. \bibinfo{publisher}{Association for Computing Machinery}, \bibinfo{address}{New York, NY, USA}, \bibinfo{pages}{2291–2302}.
\newblock
\showISBNx{9781450392211}


\bibitem[Vaswani et~al\mbox{.}(2023)]%
        {Transformer}
\bibfield{author}{\bibinfo{person}{Ashish Vaswani}, \bibinfo{person}{Noam Shazeer}, \bibinfo{person}{Niki Parmar}, \bibinfo{person}{Jakob Uszkoreit}, \bibinfo{person}{Llion Jones}, \bibinfo{person}{Aidan~N. Gomez}, \bibinfo{person}{Lukasz Kaiser}, {and} \bibinfo{person}{Illia Polosukhin}.} \bibinfo{year}{2023}\natexlab{}.
\newblock \bibinfo{title}{Attention Is All You Need}.
\newblock
\newblock
\showeprint[arxiv]{1706.03762}~[cs.CL]


\bibitem[Wang et~al\mbox{.}(2023b)]%
        {LLM4agent_survey}
\bibfield{author}{\bibinfo{person}{Lei Wang}, \bibinfo{person}{Chen Ma}, \bibinfo{person}{Xueyang Feng}, \bibinfo{person}{Zeyu Zhang}, \bibinfo{person}{Hao Yang}, \bibinfo{person}{Jingsen Zhang}, \bibinfo{person}{Zhiyuan Chen}, \bibinfo{person}{Jiakai Tang}, \bibinfo{person}{Xu Chen}, \bibinfo{person}{Yankai Lin}, \bibinfo{person}{Wayne~Xin Zhao}, \bibinfo{person}{Zhewei Wei}, {and} \bibinfo{person}{Ji-Rong Wen}.} \bibinfo{year}{2023}\natexlab{b}.
\newblock \bibinfo{title}{A Survey on Large Language Model based Autonomous Agents}.
\newblock
\newblock
\showeprint[arxiv]{2308.11432}~[cs.AI]


\bibitem[Wang et~al\mbox{.}(2023a)]%
        {wang2023describe}
\bibfield{author}{\bibinfo{person}{Zihao Wang}, \bibinfo{person}{Shaofei Cai}, \bibinfo{person}{Guanzhou Chen}, \bibinfo{person}{Anji Liu}, \bibinfo{person}{Xiaojian Ma}, {and} \bibinfo{person}{Yitao Liang}.} \bibinfo{year}{2023}\natexlab{a}.
\newblock \bibinfo{title}{Describe, Explain, Plan and Select: Interactive Planning with Large Language Models Enables Open-World Multi-Task Agents}.
\newblock
\newblock
\showeprint[arxiv]{2302.01560}~[cs.AI]


\bibitem[Wei et~al\mbox{.}(2022)]%
        {wei2022emergent}
\bibfield{author}{\bibinfo{person}{Jason Wei}, \bibinfo{person}{Yi Tay}, \bibinfo{person}{Rishi Bommasani}, \bibinfo{person}{Colin Raffel}, \bibinfo{person}{Barret Zoph}, \bibinfo{person}{Sebastian Borgeaud}, \bibinfo{person}{Dani Yogatama}, \bibinfo{person}{Maarten Bosma}, \bibinfo{person}{Denny Zhou}, \bibinfo{person}{Donald Metzler}, \bibinfo{person}{Ed~H. Chi}, \bibinfo{person}{Tatsunori Hashimoto}, \bibinfo{person}{Oriol Vinyals}, \bibinfo{person}{Percy Liang}, \bibinfo{person}{Jeff Dean}, {and} \bibinfo{person}{William Fedus}.} \bibinfo{year}{2022}\natexlab{}.
\newblock \bibinfo{title}{Emergent Abilities of Large Language Models}.
\newblock
\newblock
\showeprint[arxiv]{2206.07682}~[cs.CL]


\bibitem[Wei et~al\mbox{.}(2023)]%
        {CoT}
\bibfield{author}{\bibinfo{person}{Jason Wei}, \bibinfo{person}{Xuezhi Wang}, \bibinfo{person}{Dale Schuurmans}, \bibinfo{person}{Maarten Bosma}, \bibinfo{person}{Brian Ichter}, \bibinfo{person}{Fei Xia}, \bibinfo{person}{Ed Chi}, \bibinfo{person}{Quoc Le}, {and} \bibinfo{person}{Denny Zhou}.} \bibinfo{year}{2023}\natexlab{}.
\newblock \bibinfo{title}{Chain-of-Thought Prompting Elicits Reasoning in Large Language Models}.
\newblock
\newblock
\showeprint[arxiv]{2201.11903}~[cs.CL]


\bibitem[Weyns(2020)]%
        {SAS_Intro}
\bibfield{author}{\bibinfo{person}{Danny Weyns}.} \bibinfo{year}{2020}\natexlab{}.
\newblock \bibinfo{booktitle}{\emph{An Introduction to Self-adaptive Systems : A Contemporary Software Engineering Perspective}}.
\newblock \bibinfo{publisher}{Wiley-IEEE Computer Society Pr}.
\newblock
\showISBNx{9978-1-119-57494-1}


\bibitem[Weyns et~al\mbox{.}(2023)]%
        {10.1145/3589227}
\bibfield{author}{\bibinfo{person}{Danny Weyns} {et~al\mbox{.}}} \bibinfo{year}{2023}\natexlab{}.
\newblock \showarticletitle{Self-Adaptation in Industry: A Survey}.
\newblock \bibinfo{journal}{\emph{ACM Trans. Auton. Adapt. Syst.}} \bibinfo{volume}{18}, \bibinfo{number}{2} (\bibinfo{year}{2023}), \bibinfo{numpages}{44}~pages.
\newblock
\showISSN{1556-4665}


\bibitem[Weyns et~al\mbox{.}(2013)]%
        {DSAS_pattern}
\bibfield{author}{\bibinfo{person}{Danny Weyns}, \bibinfo{person}{Bradley Schmerl}, \bibinfo{person}{Vincenzo Grassi}, \bibinfo{person}{Sam Malek}, \bibinfo{person}{Raffaela Mirandola}, \bibinfo{person}{Christian Prehofer}, \bibinfo{person}{Jochen Wuttke}, \bibinfo{person}{Jesper Andersson}, \bibinfo{person}{Holger Giese}, {and} \bibinfo{person}{Karl~M. G{\"o}schka}.} \bibinfo{year}{2013}\natexlab{}.
\newblock \bibinfo{booktitle}{\emph{On Patterns for Decentralized Control in Self-Adaptive Systems}}.
\newblock \bibinfo{publisher}{Springer Berlin Heidelberg}, \bibinfo{address}{Berlin, Heidelberg}, \bibinfo{pages}{76--107}.
\newblock
\showISBNx{978-3-642-35813-5}
\urldef\tempurl%
\url{https://doi.org/10.1007/978-3-642-35813-5_4}
\showDOI{\tempurl}


\bibitem[White et~al\mbox{.}(2023)]%
        {LLM_prototyping}
\bibfield{author}{\bibinfo{person}{Jules White}, \bibinfo{person}{Sam Hays}, \bibinfo{person}{Quchen Fu}, \bibinfo{person}{Jesse Spencer-Smith}, {and} \bibinfo{person}{Douglas~C. Schmidt}.} \bibinfo{year}{2023}\natexlab{}.
\newblock \bibinfo{title}{ChatGPT Prompt Patterns for Improving Code Quality, Refactoring, Requirements Elicitation, and Software Design}.
\newblock
\newblock
\showeprint[arxiv]{2303.07839}~[cs.SE]


\bibitem[Wu et~al\mbox{.}(2023a)]%
        {TidyBot}
\bibfield{author}{\bibinfo{person}{Jimmy Wu}, \bibinfo{person}{Rika Antonova}, \bibinfo{person}{Adam Kan}, \bibinfo{person}{Marion Lepert}, \bibinfo{person}{Andy Zeng}, \bibinfo{person}{Shuran Song}, \bibinfo{person}{Jeannette Bohg}, \bibinfo{person}{Szymon Rusinkiewicz}, {and} \bibinfo{person}{Thomas Funkhouser}.} \bibinfo{year}{2023}\natexlab{a}.
\newblock \showarticletitle{TidyBot: personalized robot assistance with large language models}.
\newblock \bibinfo{journal}{\emph{Autonomous Robots}} (\bibinfo{year}{2023}).
\newblock
\showISSN{1573-7527}


\bibitem[Wu et~al\mbox{.}(2022)]%
        {AIChains}
\bibfield{author}{\bibinfo{person}{Tongshuang Wu}, \bibinfo{person}{Michael Terry}, {and} \bibinfo{person}{Carrie~Jun Cai}.} \bibinfo{year}{2022}\natexlab{}.
\newblock \showarticletitle{AI Chains: Transparent and Controllable Human-AI Interaction by Chaining Large Language Model Prompts}. In \bibinfo{booktitle}{\emph{Proceedings of the 2022 CHI Conference on Human Factors in Computing Systems}} \emph{(\bibinfo{series}{CHI '22})}. Article \bibinfo{articleno}{385}, \bibinfo{numpages}{22}~pages.
\newblock
\showISBNx{9781450391573}


\bibitem[Wu et~al\mbox{.}(2023b)]%
        {TaPA}
\bibfield{author}{\bibinfo{person}{Zhenyu Wu}, \bibinfo{person}{Ziwei Wang}, \bibinfo{person}{Xiuwei Xu}, \bibinfo{person}{Jiwen Lu}, {and} \bibinfo{person}{Haibin Yan}.} \bibinfo{year}{2023}\natexlab{b}.
\newblock \bibinfo{title}{Embodied Task Planning with Large Language Models}.
\newblock
\newblock
\showeprint[arxiv]{2307.01848}~[cs.CV]


\bibitem[Xi et~al\mbox{.}(2023)]%
        {LLM4agent_2}
\bibfield{author}{\bibinfo{person}{Zhiheng Xi}, \bibinfo{person}{Wenxiang Chen}, \bibinfo{person}{Xin Guo}, \bibinfo{person}{Wei He}, \bibinfo{person}{Yiwen Ding}, \bibinfo{person}{Boyang Hong}, \bibinfo{person}{Ming Zhang}, \bibinfo{person}{Junzhe Wang}, \bibinfo{person}{Senjie Jin}, \bibinfo{person}{Enyu Zhou}, \bibinfo{person}{Rui Zheng}, \bibinfo{person}{Xiaoran Fan}, \bibinfo{person}{Xiao Wang}, \bibinfo{person}{Limao Xiong}, \bibinfo{person}{Yuhao Zhou}, \bibinfo{person}{Weiran Wang}, \bibinfo{person}{Changhao Jiang}, \bibinfo{person}{Yicheng Zou}, \bibinfo{person}{Xiangyang Liu}, \bibinfo{person}{Zhangyue Yin}, \bibinfo{person}{Shihan Dou}, \bibinfo{person}{Rongxiang Weng}, \bibinfo{person}{Wensen Cheng}, \bibinfo{person}{Qi Zhang}, \bibinfo{person}{Wenjuan Qin}, \bibinfo{person}{Yongyan Zheng}, \bibinfo{person}{Xipeng Qiu}, \bibinfo{person}{Xuanjing Huang}, {and} \bibinfo{person}{Tao Gui}.} \bibinfo{year}{2023}\natexlab{}.
\newblock \bibinfo{title}{The Rise and Potential of Large Language Model Based Agents: A Survey}.
\newblock
\newblock
\showeprint[arxiv]{2309.07864}~[cs.AI]


\bibitem[Xia et~al\mbox{.}(2023)]%
        {LLM_program_repair}
\bibfield{author}{\bibinfo{person}{Chunqiu~Steven Xia}, \bibinfo{person}{Yuxiang Wei}, {and} \bibinfo{person}{Lingming Zhang}.} \bibinfo{year}{2023}\natexlab{}.
\newblock \showarticletitle{Automated Program Repair in the Era of Large Pre-trained Language Models}. In \bibinfo{booktitle}{\emph{2023 IEEE/ACM 45th International Conference on Software Engineering (ICSE)}}. \bibinfo{pages}{1482--1494}.
\newblock


\bibitem[Xu et~al\mbox{.}(2023)]%
        {xu2023Werewolf}
\bibfield{author}{\bibinfo{person}{Zelai Xu}, \bibinfo{person}{Chao Yu}, \bibinfo{person}{Fei Fang}, \bibinfo{person}{Yu Wang}, {and} \bibinfo{person}{Yi Wu}.} \bibinfo{year}{2023}\natexlab{}.
\newblock \bibinfo{title}{Language Agents with Reinforcement Learning for Strategic Play in the Werewolf Game}.
\newblock
\newblock
\showeprint[arxiv]{2310.18940}~[cs.AI]


\bibitem[Yang et~al\mbox{.}(2023)]%
        {LLM4AD_survey}
\bibfield{author}{\bibinfo{person}{Zhenjie Yang}, \bibinfo{person}{Xiaosong Jia}, \bibinfo{person}{Hongyang Li}, {and} \bibinfo{person}{Junchi Yan}.} \bibinfo{year}{2023}\natexlab{}.
\newblock \bibinfo{title}{A Survey of Large Language Models for Autonomous Driving}.
\newblock
\newblock
\showeprint[arxiv]{2311.01043}~[cs.AI]


\bibitem[Zeng et~al\mbox{.}(2023)]%
        {LLM4robotics}
\bibfield{author}{\bibinfo{person}{Fanlong Zeng}, \bibinfo{person}{Wensheng Gan}, \bibinfo{person}{Yongheng Wang}, \bibinfo{person}{Ning Liu}, {and} \bibinfo{person}{Philip~S. Yu}.} \bibinfo{year}{2023}\natexlab{}.
\newblock \bibinfo{title}{Large Language Models for Robotics: A Survey}.
\newblock
\newblock
\showeprint[arxiv]{2311.07226}~[cs.RO]


\bibitem[Zhang et~al\mbox{.}(2023a)]%
        {zhang2023controlling}
\bibfield{author}{\bibinfo{person}{Bin Zhang}, \bibinfo{person}{Hangyu Mao}, \bibinfo{person}{Jingqing Ruan}, \bibinfo{person}{Ying Wen}, \bibinfo{person}{Yang Li}, \bibinfo{person}{Shao Zhang}, \bibinfo{person}{Zhiwei Xu}, \bibinfo{person}{Dapeng Li}, \bibinfo{person}{Ziyue Li}, \bibinfo{person}{Rui Zhao}, \bibinfo{person}{Lijuan Li}, {and} \bibinfo{person}{Guoliang Fan}.} \bibinfo{year}{2023}\natexlab{a}.
\newblock \bibinfo{title}{Controlling Large Language Model-based Agents for Large-Scale Decision-Making: An Actor-Critic Approach}.
\newblock
\newblock
\showeprint[arxiv]{2311.13884}~[cs.AI]


\bibitem[Zhang et~al\mbox{.}(2023b)]%
        {zhang2023toolcoder}
\bibfield{author}{\bibinfo{person}{Kechi Zhang}, \bibinfo{person}{Huangzhao Zhang}, \bibinfo{person}{Ge Li}, \bibinfo{person}{Jia Li}, \bibinfo{person}{Zhuo Li}, {and} \bibinfo{person}{Zhi Jin}.} \bibinfo{year}{2023}\natexlab{b}.
\newblock \bibinfo{title}{ToolCoder: Teach Code Generation Models to use API search tools}.
\newblock
\newblock
\showeprint[arxiv]{2305.04032}~[cs.SE]


\bibitem[Zhao et~al\mbox{.}(2023)]%
        {LLM_survey}
\bibfield{author}{\bibinfo{person}{Wayne~Xin Zhao}, \bibinfo{person}{Kun Zhou}, \bibinfo{person}{Junyi Li}, \bibinfo{person}{Tianyi Tang}, \bibinfo{person}{Xiaolei Wang}, \bibinfo{person}{Yupeng Hou}, \bibinfo{person}{Yingqian Min}, \bibinfo{person}{Beichen Zhang}, \bibinfo{person}{Junjie Zhang}, \bibinfo{person}{Zican Dong}, \bibinfo{person}{Yifan Du}, \bibinfo{person}{Chen Yang}, \bibinfo{person}{Yushuo Chen}, \bibinfo{person}{Zhipeng Chen}, \bibinfo{person}{Jinhao Jiang}, \bibinfo{person}{Ruiyang Ren}, \bibinfo{person}{Yifan Li}, \bibinfo{person}{Xinyu Tang}, \bibinfo{person}{Zikang Liu}, \bibinfo{person}{Peiyu Liu}, \bibinfo{person}{Jian-Yun Nie}, {and} \bibinfo{person}{Ji-Rong Wen}.} \bibinfo{year}{2023}\natexlab{}.
\newblock \bibinfo{title}{A Survey of Large Language Models}.
\newblock
\newblock
\showeprint[arxiv]{2303.18223}~[cs.CL]


\bibitem[Zhou et~al\mbox{.}(2023)]%
        {zhou2023leasttomost}
\bibfield{author}{\bibinfo{person}{Denny Zhou}, \bibinfo{person}{Nathanael Schärli}, \bibinfo{person}{Le Hou}, \bibinfo{person}{Jason Wei}, \bibinfo{person}{Nathan Scales}, \bibinfo{person}{Xuezhi Wang}, \bibinfo{person}{Dale Schuurmans}, \bibinfo{person}{Claire Cui}, \bibinfo{person}{Olivier Bousquet}, \bibinfo{person}{Quoc Le}, {and} \bibinfo{person}{Ed Chi}.} \bibinfo{year}{2023}\natexlab{}.
\newblock \bibinfo{title}{Least-to-Most Prompting Enables Complex Reasoning in Large Language Models}.
\newblock
\newblock
\showeprint[arxiv]{2205.10625}~[cs.AI]


\bibitem[Zhu et~al\mbox{.}(2023)]%
        {zhu2023ghost}
\bibfield{author}{\bibinfo{person}{Xizhou Zhu}, \bibinfo{person}{Yuntao Chen}, \bibinfo{person}{Hao Tian}, \bibinfo{person}{Chenxin Tao}, \bibinfo{person}{Weijie Su}, \bibinfo{person}{Chenyu Yang}, \bibinfo{person}{Gao Huang}, \bibinfo{person}{Bin Li}, \bibinfo{person}{Lewei Lu}, \bibinfo{person}{Xiaogang Wang}, \bibinfo{person}{Yu Qiao}, \bibinfo{person}{Zhaoxiang Zhang}, {and} \bibinfo{person}{Jifeng Dai}.} \bibinfo{year}{2023}\natexlab{}.
\newblock \bibinfo{title}{Ghost in the Minecraft: Generally Capable Agents for Open-World Environments via Large Language Models with Text-based Knowledge and Memory}.
\newblock
\newblock
\showeprint[arxiv]{2305.17144}~[cs.AI]


\bibitem[Zohar et~al\mbox{.}(2023)]%
        {zohar2023lovm}
\bibfield{author}{\bibinfo{person}{Orr Zohar}, \bibinfo{person}{Shih-Cheng Huang}, \bibinfo{person}{Kuan-Chieh Wang}, {and} \bibinfo{person}{Serena Yeung}.} \bibinfo{year}{2023}\natexlab{}.
\newblock \bibinfo{title}{LOVM: Language-Only Vision Model Selection}.
\newblock
\newblock
\showeprint[arxiv]{2306.08893}~[cs.CV]


\end{thebibliography}

\end{document}